\documentclass[a4paper]{article}
\usepackage[a4paper,top=3cm,bottom=2cm,left=3cm,right=3cm,marginparwidth=1.75cm]{geometry}
\usepackage{amsfonts}
\usepackage{bm}
\usepackage{extarrows}
\usepackage{amssymb}
\usepackage{color}
\usepackage[all]{xy}
\usepackage{graphicx}
\usepackage{braket}
\usepackage{amsmath}
\usepackage{appendix}
\usepackage{graphicx}
\usepackage[colorinlistoftodos]{todonotes}
\usepackage{amsthm}
\usepackage{mathrsfs}
\usepackage{amssymb}
\usepackage{accents}
\usepackage{extarrows}
\usepackage{makecell}
\usepackage{authblk}
\usepackage{url}
\usepackage{hyperref}
\usepackage{cite}
\usepackage{eufrak}
\hypersetup{colorlinks,
            citecolor=black,
            linkcolor=black,
            urlcolor=green,
            pdftex}

\usepackage[english]{babel}
\usepackage[utf8x]{inputenc}
\usepackage[T1]{fontenc}

\usepackage[normalem]{ulem}

\title{Geometric parametrization of $SO(D+1)$ phase space of all dimensional loop quantum gravity}
\author[1,2]{Gaoping Long \footnote{201731140005@mail.bnu.edu.cn}}
\author[2]{Chun-Yen Lin \footnote{cynlin@ucdavis.edu}\thanks{corresponding author}}

\affil[1]{Department of Physics, South China University of Technology, Guangzhou 510641, China}
\affil[2]{Department of Physics, Beijing Normal University, Beijing 100875, China}

\date{}

\begin{document}

\maketitle

\begin{abstract}

To clarify the geometric information encoded in the $SO(D+1)$ spin-network states for the higher dimensional loop quantum gravity, we generalize the twisted-geometry parametrization of the $SU(2)$ phase space for $(1+3)$ dimensional loop quantum gravity to that of the $SO(D+1)$ phase space for the all-dimensional case. The Poisson structure in terms of the twisted geometric variables suggests a new gauge reduction procedure, with respect to the discretized Gaussian and simplicity constraints governing the kinematics of the theory. Endowed with the geometric meaning via the parametrization, our reduction procedure serves to identify proper gauge freedom associated with the anomalous discretized simplicity constraints and subsequently leads to the desired classical state space of the (twisted) discrete ADM data.

\end{abstract}

\section{Introduction}

Loop quantum gravity (LQG) \cite{Ashtekar2012Background}\cite{Han2005FUNDAMENTAL}\cite{thiemann2007modern}\cite{rovelli2007quantum} as a candidate theory of quantum gravity provides a possibility of deriving general relativity (GR) from the foundation of plank-scale quantum geometry. Thus the theory, in a broader context, provides a concrete platform for exploring the relation between the continuum classical GR variables and the discretized geometric quantum data, such as those of the twistor theory and Regge calculus \cite{Freidel_2010}\cite{freidel2010twisted}. On the other hand, it has been realized that the correspondence between the field-geometric and the quantum data is far beyond the issue of merely taking the continuum limits. This is due to the fact that canonical GR is governed by a constraint system, and the correspondence may be fully revealed only for the physical degrees of freedom--- with all the constraints properly imposed in the quantum theory. From the opposite direction of this view, the concrete goal of recovering the familiar ADM data from LQG may provide useful instructions in tackling the abstract problems of quantum constraint reductions in the theory.

A series of illuminating analysis in this direction has been carried out in the case of the $SU(2)$ formulation of $(1+3)$-dimensional loop quantum gravity. Based on the Ashtekar formulation of canonical GR using the $SU(2)$ desitized triad and connection conjugate variables, LQG in this formulation has a kinematic Hilbert space spanned by the spin-network states, each of which is given by a network of the connection holonomies, with each edge of the graph of the network colored by a specific $SU(2)$ representation, and each of the vertices colored by an intertwiner specifying a coupling among the neighboring $SU(2)$ representations. Under the well-defined flux-holonomy geometric operators, the $SU(2)$ representations indicate the quanta of the triad-fluxes as the area elements dual to the graph's edges, while the intertwiners indicate the intersection angles amongst these triad-fluxes at the vertices. This discretized distribution of the $2$-dimensional spatial area elements with the intersection angles leads to a specific notion of quantum geometry that is the foundation of LQG. The classical constraints-- the scalar, vector and $SU(2)$ Gauss constraints---can be represented via the flux-holonomy operators for the quantum theory. It has been shown that the imposition of the quantum Gauss constraints on the coherent spin-network states gives rise to a proper semi-classical symplectic reduction, in the holonomy-flux phase of the discretized Ashtekar formulation on the given graph. Remarkably, in the reduced state space, the coherent spin-network states satisfying the quantum Gauss constraints not only describe the intrinsic spatial geometry built from the polytope-cells dual to the network \cite{Freidel_2010}\cite{bianchi2011polyhedra}\cite{Conrady_2009}, but also carry precisely the right data to specify the extrinsic curvature of the hypersurface made of the polytopes. \cite{freidel2010twisted}\cite{Bianchi:2009ky}. Through this first stage of the semi-classical gauge reduction, a notion of kinematic ADM data may thus appear in the discrete form of Regge geometry, upon which the further reductions with the momentum and scalar constraints should to be carried out. The quantum vector and scalar constraints take much more complicated forms in the flux-holonomy operators, and unlike the quantum Gauss constraint, their anomalous algebra is no longer of first class. With the quantum anomaly hindering the standard Dirac procedure mirroring the classical gauge reduction, the treatment of these loop-quantized ADM constraints remains a crucial challenge for LQG tackled by many ongoing projects.

As we introduced above, loop quantum gravity is first constructed as a quantum theory of GR in four dimensional spacetime. Nevertheless, with the various classical and quantum gravity theories in higher-dimensional spacetimes (i.e., Kaluza-Klein theory, super string theories) showing remarkable potentials in unifying gravity and other fundamental interactions, it has been recognized that the framework of arbitrary dimensional loop quantum gravity may serve as a novel approach toward the higher-dimensional ideas of unification, upon the background-independent and non-perturbative construction of the discretized quantum geometry. Pioneered by Bodendorfer, Thiemann and Thurn \cite{bodendorfer2013newi}\cite{bodendorfer2013newiii}\cite{bodendorfer2013implementation}, the loop quantization approach for general relativity in all dimensions has been developed. In the context of the higher dimensional loop quantum gravity, the challenge of loop quantum anomaly already exists at the kinematic level before the accounts of the quantum ADM constraints; though, here it is in a simpler form for us to develop concrete insights and solutions to the problem. In detail, the all dimensional LQG  is based on the universal Ashtekar formulation of $(1+D)$ dimensional general relativity in the form of the $SO(D+1)$ Yang-Mills theory, with the kinematic phase space coordinatized by the canonical pairs $(A_{aIJ},\pi^{bKL})$, consisting of the spatial $SO(D+1)$ connection fields $A_{aIJ}$ and the vector fields $\pi^{bKL}$. In this formulation, the theory is governed by the first class system of the $SO(D+1)$ Gauss constraints, the $(D+1)$-dimensional ADM constraints and the additional constraints called the simplicity constraints. Taking the form $S^{ab}_{IJKL}:=\pi^{a[IJ}\pi^{|b|KL]}$, the simplicity constraints generate extra gauge symmetries in the $SO(D+1)$ Yang-Mills phase space. It is known that the phase space correctly reduces to the familiar ADM phase space after the symplectic reductions with respected to the Gauss and simplicity constraints. Similar to the case of the $SU(2)$ formulation, the loop quantization of the $SO(D+1)$ formulation leads to the spin-network states of the $SO(D+1)$ holonomies carrying the quanta of the flux operators representing the flux of $\pi^{bKL}$ over a $(D-1)$ surface. Following the previous experience, one may attempt to look for the all-dimensional Regge ADM data encoded in the $SO(D+1)$ spin-network states, through a gauge reduction procedure with respect to both the quantum $SO(D+1)$ Gaussian constraints and the quantum simplicity constraints.

This is where the challenge arises--- the standard quantum simplicity constraints in LQG carry serious quantum anomaly. As a result of the loop quantization, the abelian algebra of the classical simplicity constraints becomes the deformed algebra of the quantum simplicity constraints that is not even close \cite{bodendorfer2013implementation}. As an important consequence, the transformations generated by these anomalous quantum simplicity constraints can happen between states supposed to be physically distinct in terms of the semiclassical limits. Strong impositions of the quantum simplicity constraints thus lead to over-constrained physical states unable to reproduce the semi classical degrees of freedom.
In a closer look, the quantum simplicity constraints in LQG consist of two types of local constraints due to the network discretization--- the edge-simplicity constraints and the vertex-simplicity constraints. Importantly, the algebra anomaly happens only amongst the vertex-simplicity constraints, while the edge-simplicity constraints remain anomaly free in the sense of having a weakly abelian algebra. Previously, we have proposed a new method \cite{Gaoping2019coherentintertwiner} of weakly imposing the anomalous vertex-simplicity constraints for the vanishing expectation values and minimal quantum fluctuations, upon a special class of states in the space of the $SO(D+1)$-invariant spin-network states satisfying the quantum Gauss constraints. With their edges labeled by only the simple representations and their vertices by specific coherent states of intertwiners, this class of states strongly satisfy the quantum edge-simplicity constraints and are sharply peaked for the flux operators. We found that, among this class of states, each weak solution of the vertex-simplicity constraints describes a set of quantum $D$-dimensional polytopes dual to the vertices of its graph. Also, in large quantum-number limits, these weak solutions indeed recover all the degrees of freedom in the classical $D$-dimensional polytopes, which may be assembled to describe all possible states of quantum spatial geometry. Concerning the proper gauge-reduction procedure, this remarkable result suggests that, in the space of strong solutions to the first class system of the quantum Gauss constraints and edge-simplicity constraints, the vertex-simplicity constraints should serve as additional constraints---unrelated to the quantum gauge symmetries---and select the special gauge-invariant states capable of giving the desired quantum discrete spatial intrinsic geometry.

Toward the full ADM data of a hypersurface, our remaining task is to show how the polytopes at the various vertices are correlated, so that both of the intrinsic and extrinsic geometry can be recovered. Clearly, this task would be completing the other half gauge-reduction procedure following our strategy above: identifying the proper gauge orbits associated with the quantum simplicity constraints and finding the geometrical interpretation to the invariant degrees of freedom. It is known that the classical simplicity constraints transform only the pure-gauge components of the $SO(D+1)$ Ashtekar connection $A_{aIJ}$, while leaving the vector fields $\pi^{bKL}$ invariant. We will demonstrate that this picture could emerge in the LQG flux-holonomy phase space associated with a graph, following the other half of our reduction procedure at the classical and discrete level under a satisfying geometric interpretation, in spite of the anomaly of quantum simplicity constraint already appearing in this classical and discrete formulation. In the crucial step for establishing such an interpretation, we will generalize the existing twisted-geometry parametrization for the $SU(2)$ flux-holonomy phase space, into that for the $SO(D+1)$ setting. These new geometric coordinates for the phase space, along with their well-formulated expressions for the symplectic structure, enable the full analysis of the gauge reductions in the language of the twisted geometry.

Our result shows that, the discretized classical Gaussian, edge-simplicity and vertex-simplicity constraints catching the anomaly of quantum vertex simplicity constraint define a constraint surface in the discrete phase space of all dimensional LQG, and the kinematic physical degrees of freedom parameterized by the generalized hypersurface twisted-geometry are given by the gauge orbits in the constraint surface generated by the first class system of discrete Gaussian and edge-simplicity constraints. In particular, we find the orbits of the edge-simplicity constraints to be along the angle variables of the twisted geometry, which indeed represent the smeared form of the pure-gauge components of the Ashtekar connection in the continuous theory. Finally, the complete ADM data of a Regge hypersurface can be identified as the degrees of freedom of the reduced generalized twisted geometry space, under an additional condition called the shape matching condition.

In our brief review of the classical Ashtekar formulation of all dimensional GR in Section 2, we will also introduce the flux-holonomy phase space for the discretized formulation with the anomalous vertex simplicity constraints. In Section 3 and Section 4 we will introduce the twisted-geometry parametrization for the $SO(D+1)$ phase space, and analyze the Poisson structures among the new geometric parametrization variables and the discretized simplicity constraints. Finally in Section 5 we will combine the obtained gauge transformations with the geometric interpretations, and formalize the gauge reduction procedure that leads to the desired ADM data. We will then conclude with the outlook for the possible next steps of the future research.

\section{Phase space of all dimensional loop quantum gravity and simplicity constraint}

The classical Ashtekar formulation of general relativity with arbitrary spacetime dimensionality of $(D+1)$ has been developed by Bodendofer, Thiemann and Thurn in \cite{bodendorfer2013newi}. The continuum connection phase space of the theory is coordinatized by a $so(D+1)$ valued canonical pair $(A_{aIJ}, \pi^{bKL})$ with the non-trivial Poisson brackets
\begin{equation}\label{Poisson1}
\{A_{aIJ}(x), \pi^{bKL}(y)\}=2\kappa\beta\delta_a^b\delta_{[I}^K\delta_{J]}^L\delta^{(D)}(x-y),
\end{equation}
where $\beta$ is the Barbero-Immirzi parameter and $\kappa$ is the gravitational constant. It is known that this phase space correctly reduces to the familiar ADM phase space after the standard sympletic reduction procedure with respect to the first-class constraint system of the Gauss
constraints $\mathcal{G}^{IJ}\approx0$ and simplicity constraints $S^{ab}_{IJKL}:=\pi^{a[IJ}\pi^{|b|KL]}\approx0$. Specifically, the spatial metric $q_{ab}$ is given by $q_{ab}=e_{aI}e_b^I$, where $e_a^I$ is a D-bein field parametrizing the simplicity constraint solutions in the form $\pi^{aIJ}=2\sqrt{q}{\mathcal{N}}^{[I}e^{|a|J]}$ together with a chosen field ${\mathcal{N}}^I$ satisfying ${\mathcal{N}}^I{\mathcal{N}}_I=1$ and ${\mathcal{N}}^Ie_{aI}=0$. The densitized extrinsic curvature is given by $\tilde{K}_a^{\ b}=K_{aIJ}\pi^{bIJ}$ where $K_{aIJ}$ is the component of $A_{aIJ}$ under the splitting
\begin{equation}\label{splitA}
A_{aIJ}\equiv\Gamma_{aIJ}(e)+\beta K_{aIJ}
\end{equation}
on simplicity constraint surface, where $\Gamma_{aIJ}(e)$ is the unique torsionless spin connection compatible with the D-bein $e_{aI}$.

Let us look into the simplicity constraints from the perspectives of the corresponding reductions. First, the solutions $\pi^{aIJ}=2\sqrt{q}{\mathcal{N}}^{[I}e^{|a|J]}$ to the quadratic simplicity constraints introduced above defines the constraint surface of the simplicity constraints. It is easy to check that the infinitesimal gauge transformations induced by simplicity constraints are given by
\begin{equation}\label{gaugetranssim}
\delta K_{c}^{PQ}=\{\int_{\sigma}d^Dxf_{ab}^{IJKL}\pi^a_{[IJ}\pi^b_{KL]}(x), K_{c}^{PQ}(y)\}=4\kappa\beta f_{cb}^{[PQKL]}\pi^b_{KL}(y).
\end{equation}
On the simplicity constraint surface we have $\pi^{aIJ}=2\sqrt{q}{\mathcal{N}}^{[I}e^{|a|J]}$ and thus
$\delta K_{c}^{IJ}{\mathcal{N}}_I=0$. Therefore, introducing the decomposition of $K_{aIJ}$ as
\begin{equation}
K_{aIJ}\equiv 2{\mathcal{N}}_{[I}K_{|a|J]}+\bar{K}_{aIJ},
\end{equation}
where $\bar{K}_{aIJ}:=\bar{\eta}_I^K\bar{\eta}_J^LK_{aKL}$ with $\bar{\eta}^I_J=\delta^I_J-{\mathcal{N}}^I {\mathcal{N}}_J$ and $\bar{K}_{aIJ}{\mathcal{N}}^I=0$, we immediately see that the longitudinal components $\bar{K}_{aIJ}$ parametrize the gauge redundancy, while the transverse components $2{\mathcal{N}}_{[I}K_{|a|J]}$ are gauge invariant based on the  transformations given in (\ref{gaugetranssim}). From the expressions for the ADM variables $\tilde{\tilde{q}}^{ab}=\frac{1}{2}\pi^{aIJ}\pi^b_{IJ}$ and $\tilde{K}_a^{\ b}=K_{aIJ}\pi^{bIJ}$, it is easy to see that these variables are indeed invariant under the gauge transformations by the simplicity constraints. Through the sympletic gauge-reduction procedure, the simplicity constraints thus eliminate the two sets of degrees of freedom-- setting $\bar{\pi}^{aIJ}:=\pi^{aIJ}-2\sqrt{q}n^{[I}e^{|a|J]}\approx0$ by the restriction to the constraint surface and removing the pure-gauge components $\bar{K}_{aIJ}:=\bar{\eta}_I^K\bar{\eta}_J^LK_{aKL}$.

The foundation leading to the quantum geometry of loop quantum gravity is the use of the spatially smeared variables-- the D-bein fluxes over surfaces and connection holonomies over paths-- for the conjugate pairs of elementary variables. The quantization of the flux-holonomy algebra leads to the space of spin-network states mentioned above, spanned by the basis states of holonomy networks each labeled by a graph with the representation and intertwiner colorings. We will focus on the holonomies and fluxes based on one specific graph for the following. The edges of the given graph naturally provide the set of paths for a fixed set of holonomies, and the cell decomposition dual to the graph provides the set of (D-1)-faces specifying a fixed set of fluxes. In this setting, the holonomy over one of the edges is naturally conjugating to the flux over the face traversed by the edge, and the pairs associated with the given graph satisfy the smeared version of the algebra (\ref{Poisson1}) and form a new phase space. More precisely, given the graph $\gamma$ embedded in the spatial manifold, we consider a new algebra given by replacing $(A_{aIJ},\pi^{bKL})$ with the pairs $(g_e, X_e)\in SO(D+1)\times so(D+1)$ over all edges $e$ of $\gamma$. These pairs of variables represent the discretized version of the connection and its conjugate momentum $\pi^{aIJ}$, respectively via the holonomies $g_e=\mathcal{P}\exp\int_e A$ with $\mathcal{P}$ denoting the path-ordered product, and fluxes $X_e=\int_{e^\star}(g\pi g^{-1})^an_ad{}^{D-1}\!S$ with $e^\star$ being the dual (D-1)-dimensional face to the edge $e$, with the normal $n_a$ and infinitesimal coordinate area element $d{}^{D-1}\!S$ and $g$ is the parallel transport from one fixed vertex to the point of integration along a path adapted to the graph. Since $SO(D+1)\times so(D+1)\cong T^\ast SO(D+1)$, this new discrete phase space called the phase space of $SO(D+1)$ loop quantum gravity on a fixed graph, is a direct product of $SO(D+1)$ cotangent bundles. Finally, the complete phase space of the theory is given by taking the union over the phase spaces of all possible graphs. Just like the $SU(2)$ case, the new variables $(g_e, X_e)$ of the phase space of $SO(D+1)$ loop quantum gravity can be seen as a discretized version of the continuum phase space.

A series of studies following the original works by Freidel and Speziale show that the mentioned phase space of $SU(2)$ loop quantum gravity carries the notion of what is called the twisted geometry \cite{Freidel_2010}\cite{freidel2010twisted}, and this space can undergo a symplectic reduction with respect to the discretized Gauss constraints (associated with the quantum Gauss constraint operators), giving rise to a reduced phase space containing the discretized ADM data of a polyhedral Regge hypersurface. Based on such a foundation, our first goal is providing a generalization to the above approach for the $SO(D+1)$ formulation. This includes the generalization of the twisted-geometry parametrization for the $SO(D+1)$ phase space, which should provide a clear correspondence between the original variables $(g_e, X_e)$ and the hypersurface geometry data. Our second goal is addressing the proper treatment of the (discretized) Gaussian and simplicity constraints, following the geometric meaning of the phase space under the new parametrization. We will use the standard forms of the (discretized) Gaussian and simplicity constraints in agreement with the quantum constraints. With $X_{-e}=-g_e^{-1}X_eg_e\equiv \tilde{X}_e$, the (discretized) Gauss constraints $G_v^{IJ}\approx0$ for each vertex $v\in \gamma$ of the graph take the form
\begin{eqnarray}
\label{gaussconstr}
G_v^{IJ}=\sum_{e|s(e)=v}X_e^{IJ}+\sum_{e|t(e)=v}\tilde{X}_e^{IJ}\approx0,
\end{eqnarray}
where $s(e)$ and $t(e)$ respectively denote the source and target vertices of the oriented edge $e$. The (discretized) simplicity constraints consist of the edge-simplicity constraints $S^{IJKL}_e\approx0$ and vertex-simplicity constraints $S^{IJKL}_{v,e,e'}\approx0$ taking the forms
\begin{equation}
\label{simpconstr}
S_e^{IJKL}\equiv X^{[IJ}_e X^{KL]}_e\approx0, \ \forall e\in \gamma,\quad S_{v,e,e'}^{IJKL}\equiv X^{[IJ}_e X^{KL]}_{e'}\approx0,\ \forall e,e'\in \gamma, s(e)=s(e')=v.
\end{equation}
As we mentioned in the introduction, since the commutative set of conjugate momentum varaibles $\{\pi^{bKL}\}$ becomes non-commutative set of flux variables $\{ X^{KL}_e\}$ after the smearing, these discrete version of simplicity constraints become non-commutative and thus anomalous.

\section{Geometric parametrization of edge-simplicity constraint surface in $SO(D+1)$ phase space}

\subsection{Bivector parametrization of intrinsic geometry}

In our previous work \cite{Gaoping2019coherentintertwiner}, we have explicitly constructed specific flux-coherent states based on a chosen graph $\gamma$, which are sharply peaked in every pair of flux variables associated with the source and target points of each of the edges, while having the coloring of the edges restricted to the $SO(D+1)$ simple representations. Such restriction to the irreducible representations has been shown to strongly solve the quantum edge-simplicity constraints. Subsequently, this implies that the flux expectation values associated to the source and target points of an edge must respectively take the form $N_e V_e$ and $N_e \tilde{V}_e$, with $V_e$ and $\tilde{V}_e$ given by normalized bivectors in $\mathbb{R}^{D+1}$ and the shared norm $N_e$ is the Casimir value labeling the simple representation. We have shown that, further, the quantum Gauss and quantum vertex-simplicity constraints can be weakly imposed upon our flux-coherent states by restricting and correlating the values of $(V_e, \tilde{V}_e, N_e)$ over all the edges , and the resulted states describe the familiar discrete geometry of a set of D-polytopes dual to the graph $\gamma$, with the corresponding faces dual to the same edge having the same area. In more details, with the fixed oriented graph $\gamma$, each of our flux-coherent states has a $SO(D+1)$ simple representation $N_e\in\mathbb{R}$ to each of the edges, assigning the area of the corresponding set of dual (D-1)-dimensional surfaces; for each vertex $v\in \gamma$ connected to $n_v$ number of edges, the state is also peaked at the $n_v$ number of unit bi-vectors $V^{IJ}_e(v)$ assigning the directions of the $n_v$ number of the (D-1)-surfaces. These parameters describe a direct product space
\begin{equation}
P^{\textrm{aux}}_\gamma\equiv\times_e \mathbb{R}_e\times_v P_v, \quad P_v\equiv\times_{e:v=b(e)\ \textrm{or}\ v=t(e)} Q^e_{D-1},
\end{equation}
where $Q_{D-1}:=SO(D+1)/(SO(D-1)\times SO(2))$ is the space of unit bi-vectors $V^{IJ}_e(v)$. Note that by our assignment, each edges is labelled by $N_e$ and two unit bi-vectors. Calling $s(e)$ the source vertex and $t(e)$ the target vertex of an edge $e$, we denote the two bivectors as $V_e\equiv V^{IJ}_e=V^{IJ}_e(s(e))$ and $\tilde{V}_e\equiv \tilde{V}^{IJ}_e=V^{IJ}_e(t(e))$. We may use this notation to factorize the space as
\begin{equation}
P^{\textrm{aux}}_\gamma=\times_e P^{\textrm{aux}}_e, \quad P^{\textrm{aux}}_e= Q^e_{D-1}\times Q^e_{D-1}\times \mathbb{R}_e,
\end{equation}
and the variables associated to each edge of the graph are thus a triple $(V_e,\tilde{V}_e,N_e)$. For our flux-coherent states \cite{Gaoping2019coherentintertwiner}, the weak imposition of the quantum Gauss and vertex-simplicity constraints amounts to imposing the corresponding constraints in the space $P^{\textrm{aux}}_\gamma$. The weak imposition of the quantum Gauss constraints at a vertex $v$ implies
\begin{equation}
\label{Gauss}
C_{\vec{N}_v}\equiv \sum_{e:v=b(e)}N_eV^{IJ}_e+\sum_{e: v=t(e)}N_e\tilde{V}^{IJ}_e=0,
\end{equation}
and that of the vertex-simplicity constraints at a vertex $v$ (weakly) implies
\begin{equation}
S^{IJKL}_{v}\equiv V^{[IJ}_{e_\imath}(v)V^{KL]}_{e_\jmath}(v)=0,\quad \forall e_\imath, e_\jmath :v=e_\imath\cap e_\jmath.
\end{equation}
The variables $V^{IJ}_e(v)$ for the vertex $v$ satisfying both conditions defining the common constraint surface
\begin{equation}
\mathcal{P}^{\textrm{s.}}_{\vec{N}_v}=\{(V_{e_1}^{IJ}(v),...,V_{e_{n_v}}^{IJ}(v))\in P_v| C_{\vec{N}_v}=0,\ S^{IJKL}_{v}=0 \}
\end{equation}
must take the form of $V_{e}^{IJ}(v)=\mathcal{N}^{[I}(v) V_{e}^{J]}(v)$, where the vectors $V_{e}^{J}(v)\in \mathbb{R}^{D+1}$ for each $v$ lie in the subspace $\mathbb{R}^{D}\subset\mathbb{R}^{D+1}$ orthogonal to an unit vector $\mathcal{N}^{I}(v)\in \mathbb{R}^{D+1}$, and they satisfy the familiar $D$-dimensional Minkowski closure conditions applied through the closure constraints $ C_{\vec{N}_v}=0$. Therefore these solutions define the space of flat D-polytopes embedded in the flat space $\mathbb{R}^{D}\subset\mathbb{R}^{D+1}$, and each of the $n_v$-valent vertex $v$ can be thought of as dual to a flat convex D-polytope whose $n_v$ number of (D-1)-faces' areas and normal vectors are given by repectively $\{N_e| b(e)\ \text{or}\ t(e)=v\}$ and $\{V^{J}_e,\tilde{V}^{J}_{e'}| b(e)=v,\ \text{and}\ t(e')=v\}$ satisfying the closure conditions. Since the shape of a D-polytope is invariant under the rotations, it is useful to introduce the space $\mathfrak{P}^{\textrm{s.}}_{\vec{N}_v}$ of shapes of the D-polytopes, i.e. the space of closed normals modulo the vertex-wise $SO(D+1)$ rotations as \cite{Gaoping2019coherentintertwiner}
\begin{equation}
\mathfrak{P}^{\textrm{s.}}_{\vec{N}_v}\equiv \mathcal{P}^{\textrm{s.}}_{\vec{N}_v}/SO(D+1).
\end{equation}
Therefore, we see that the bivector variables $( N_e{V}_e, N_e\tilde{V}_e)$ carried by the flux-coherent states, when taking the onshell values of the Gauss and vertex-simplicity constraints, may give a notion of discretized spatial geometry as an assembly of the locally flat D-polytopes dual to the vertices, with the identical areas for the pairs of corresponding faces amongst neighboring polytopes.

We want to complete the description of such geometry and extend it to the extrinsic part, so that a notion of hypersurface ADM data could be identified for the relevant region of the LQG phase space. Since the $D$-polytope geometry arises only after imposing the edge-simplicity constraints,  in this context the relevant region of the LQG phase space is expected to be the edge-simplicity constraint surface. This surface, denoted as $\times_e T_s^\ast SO(D+1)_e$, is obtained from the discrete phase space $\times_e T^\ast SO(D+1)_e$ by restricting the $X_e$ to be of the bivector form of $N_e V^{IJ}_e $.

As shown in \cite{mladenov1985geometric}\cite{ii1981geometric}, the space $Q_{D-1}$ is a $2(D-1)$-dimensional phase space with the invariant Kahler form $\Omega_{N^2/2}$, and the $SO(D+1)$ orbits in $P_v$ are generated precisely by the closure constraints; therefore we may construct the $SO(D+1)$-reduced phase space given by
\begin{equation}
\label{gausssolution}
\mathfrak{P}_{\vec{N}_v}=\{(V_{e_1}^{IJ}(v),...,V_{e_{n_v}}^{IJ}(v))\in P_v| C_{\vec{N}_v}=0 \}/SO(D+1).
\end{equation}
The Poisson structure on this $n_v(\frac{D(D+1)}{2}-1-\frac{(D-2)(D-1)}{2})-D(D+1)=2n_v(D-1)-D(D+1)$-dimensional space is obtained from $\Omega_{N^2/2}$ defined in $Q_{D-1}$, via the standard symplectic reduction. However, subject to the additional vertex-simplicity constraints the space $\mathfrak{P}^{\textrm{s.}}_{\vec{N}_v}$ describing the shapes of the D-polytopes is not a phase space because the imposition of vertex-simplicity constraints clearly does not give a symplectic reduction \cite{Gaoping2019coherentintertwiner}. Just as mentioned, we will demonstrate that a true reduction can be carried out in the discrete phase space extended from $P^{\textrm{aux}}_\gamma$, which includes the information about the extrinsic curvature and contains the gauge degrees of freedom for the discretized simplicity constraints.

From now on, we denote the symplectic reduction by double quotient $\mathfrak{P}_{\vec{N}_v}=P_v//C_{\vec{N}_v}$. Considering the space $P_{\gamma}^{\textrm{aux}}$ defined for the whole graph $\gamma$, we can accordingly apply the symplectic reduction by $\mathcal{C}_\gamma=\{C_{\vec{N}_v}|v\in\gamma\}$ and impose both the closure conditions and vertex-simplicity constraints on all the vertices. The result is
\begin{equation}
\mathcal{K}_\gamma\equiv P_{\gamma}^{\textrm{aux}}//\mathcal{C}=\times_e\mathbb{R}_e\times_v \mathfrak{P}_{\vec{N}_v}, \quad \mathcal{K}_\gamma^{\textrm{s.}}\equiv\mathcal{K}_\gamma|_{S_v^{IJKL}=0, \forall v\in \gamma}=\times_e\mathbb{R}_e\times_v \mathfrak{P}^{\text{s.}}_{\vec{N}_v}.
\end{equation}

\subsection{Full parametrization}

According to our discussion above, the space $(g_e, X_e)$ is parametrized by $ (g_e, N_e V_e)$ in the constraint surface $\times_e T_s^\ast SO(D+1)_e$. However, this is not the end of the story. For describing the hypersurface geometry, it is also important to express $(g_e, V_e, \tilde{V}_e, N_e)$ in terms of the variables clearly describe the extrinsic geometry distinctly from the intrinsic geometry, so that a concrete notion of hypersurface can emerge after proper impositions of the constraints. A method for this goal has been studied in the $SU(2)$ formulation, and in the following we will explicitly work out the generalization of such method for our $SO(D+1)$ case. In both cases the key lies in extracting the an angle variables from the values $(g_e, V_e, \tilde{V}_e, N_e)$, that capture the extrinsic curvature of the hypersurface.

To extract the extrinsic data, we first identify the intrinsic geometric data completely. Recall the emerging D-polytopes dual to the vertices, that for any two D-polytope next to each across an edge, the neighboring pair of (D-1)-faces associated to an edge always have the same area. As observed in the $SU(2)$ case, when the neighboring pair of faces are subject to an additional ``shape matching" condition that they have the same shape in addition to the same area, each of these special values of $( N_e{V}_e, N_e\tilde{V}_e)$ assigns one discretized intrinsic-geometry of a D-dimensional hypersurface, given by simply gluing the neighboring identical faces of the D-polytopes. Such geometry is just the spatial Regge geometry, with the local geometry within each D-polytope flat and the curvature of the hypersurface captured in the parallel transports amongst the constituent D-polytopes. The area-matching D-polytopes dual to a graph $\gamma$ without the shape matching conditions define a more general notion of geometry, which is called twisted geometry in the existing literature.

Now let us look at the construction for the twisted geometry associated to $\gamma$ in all-dimensional case. Note that two neighboring D-polytopes prescribed by the flux variables must be rotated by a specific $SO(D+1)$ element, for their identified pair of faces to aligned in the normal directions. In this manner, the flux data can specify one $SO(D+1)$ element to every edge $e$ as the necessary operation to align and glue the pair of faces dual to $e$, and this element should rotate the inward (area-weighted) normal $-N_e\tilde{V}_e$ of the (D-1)-face for the the target D-polytope, into the outward normal $N_e{V}_e$ of the corresponding (D-1)-face for the source D-polytope. Under the generalized Regge geometry interpretation \cite{Alesci:2014aza}, these transformations across the edges carry the meaning of the Levi-Civita holonomies. We thus define the $SO(D+1)$ valued Levi-Civita holonomy $h^{\Gamma}_{e}$ for every edge $e$ as a function of the bivector variables associated to the neighborhood of $e$. Note that, by construction we have ${V}_e= -h^{\Gamma}_{e}\circ \tilde {V}_e:=-h^{\Gamma}_{e}\tilde {V}_e(h^{\Gamma}_{e})^{-1}$.

We now adopt a decomposition of the holonomy as a $SO(D+1)$ element in the following way. First, we choose once for all a fixed generator $\tau_0 \in so(D+1)$ as a reference bivector $\tau^{IJ}_0\equiv (\frac{\partial}{\partial x_1})^{[I}(\frac{\partial}{\partial x_2})^{J]}$. Then for each edge $e\subset \gamma$, we specify a special pair of differentiable $SO(D+1)$-valued functions of the bi-vector variables called the Hopf sections, denoted as $u_e( {V}_{e})$ and $\tilde{u}_e( \tilde{V}_{e})$; the Hopf sections for each edge $e$ are defined by the conditions
\begin{eqnarray}
\label{decomp1}
V_e=u_e \tau_0 u^{-1}_e\,\,,\,\, \tilde{V}_e=-\tilde{u}_e \tau_0 \tilde{u}^{-1}_e\,\,
\text{and }\,\, u_e( -{V}_{e})=u_e( {V}_{e})e^{2\pi\tau_{13}}, \tilde{u}_e( -{\tilde{V}}_{e})=\tilde{u}_e( {\tilde{V}}_{e})e^{2\pi\tau_{13}}
\end{eqnarray}
with $\tau_{13}:= (\frac{\partial}{\partial x_1})^{[I}(\frac{\partial}{\partial x_3})^{J]}$ and $e^{2\pi\tau_{13}}\tau_{0}e^{-2\pi\tau_{13}}=-\tau_{0}$. Observe that the choice for the Hopf sections is clearly non-unique, and from now on our parametrization will be given under one fixed choice of $\{u_e,\tilde{u}_e\}$ for every edge $e$, under which the Levi-Civita holonomy $h^{\Gamma}_{e}$ can be expressed in the form
\begin{equation}\label{hgamma}
h^{\Gamma}_{e}( {V}_{e'},\tilde{V}_{e'})\equiv u_e\,\, (e^{\bar{\eta}_{e}^{\mu}\, \bar\tau_\mu}\, e^{\eta_{e}\,\tau_o})\,\,\tilde{u}_e^{-1},
\end{equation}
where the $ e^{\bar{\eta}^\mu \bar\tau_\mu}$ takes value in the subgroup $SO(D-1)\subset SO(D+1)$ preserving both $\frac{\partial}{\partial x_1}$ and $\frac{\partial}{\partial x_2}$. Note that the bivector functions $\eta_{e}$ and $\bar{\eta}_{e}^{\mu}$ are well-defined via the given $h^{\Gamma}_{e}$ and the chosen Hopf sections. Accordingly, the holonomy $g_e$ assigned to edge $e$ can also be decomposed as
\begin{eqnarray}
\label{decomp2}
&&g_e(V_{e'},\tilde{V}_{e'},\xi_e,\bar{\xi}_e^\mu)\equiv u_e\,\, (e^{\bar{\xi}_e^\mu \bar\tau_\mu}\, e^{\xi_e\tau_o})\,\,\tilde{u}_e^{-1}.
\end{eqnarray}
Observe that while the $\bar{\eta}_{e}^{\mu}$ and $\eta_{e}$ are already fixed by the given $h^{\Gamma}_{e}( {V}_{e'},\tilde{V}_{e'})$ and the Hopf sections, the free variables $\bar{\xi}_e^\mu$ and $\xi_e$, which we will call the angle variables, parametrize the additional degrees of freedom in $g_e$. Moreover, we can factor out $h^{\Gamma}_{e}$ from $g_e$ through the expressions
\begin{eqnarray}
\label{decomp3}
g_e= h^{\Gamma}_{e}\,\, \left(e^{-\bar{\eta}_e^\mu \tilde{u}_{e}\!\bar\tau_\mu\!\tilde{u}^{-1}_{e}}\,e^{\bar{\xi}_e^\mu \tilde{u}_{e}\!\bar\tau_\mu\!\tilde{u}^{-1}_{e}}\, e^{-(\xi_e- \eta_e) \tilde{V}_e}\right) =\left(e^{\bar{\xi}_e^\mu {u}_{e}\!\bar\tau_\mu\!{u}^{-1}_{e}}\,e^{ -\bar{\eta}_e^\mu {u}_{e}\!\bar\tau_\mu\!{u}^{-1}_{e}}\, e^{(\xi_e- \eta_e) {V}_e}\right)\,\,h^{\Gamma}_{e},
\end{eqnarray}
where the ${u}_{e}\bar\tau_\mu{u}^{-1}_{e}$ or $\tilde{u}_{e}\bar\tau_\mu\tilde{u}^{-1}_{e}$ takes values from the subgroups $SO(D-1)\subset SO(D+1)$ respectively preserving the bivector ${V}_e$ or $\tilde{V}_e$.

Having introduced the parametrization defined by \eqref{decomp1}-\eqref{decomp3} for all points in the phase space, we now focus on the points parametrized by the special angle-bivector values with the flux data $( N_e{V}_e, N_e\tilde{V}_e)$ describing a Regge intrinsic geometry, so that we can identify the extrinsic curvature data for these states through the angle variables in the following manner. The above decomposition with the angle-bivector variables suggests a splitting of the the Ashtekar connection as $A_a=\Gamma_a+\beta K_a$ on a given graph. For that, consider the integral of $A_a=\Gamma_a+\beta K_a\in so(D+1)$ along an infinitesimal edge direction $\ell^a_e$ leading to $A_e\equiv A_a\ell^a_e$, $\Gamma_e\equiv \Gamma_a\ell^a_e$ and $K_e\equiv K_a\ell^a_e$. Clearly, we have the following obvious correspondence of
\begin{eqnarray}
\label{corres1}
g_e= e^{A_e} \,\,\,\text{and}\,\,\,h^{\Gamma}_{e}= e^{\Gamma_e}.
\end{eqnarray}
The remaining factor should account for the $K_e$. Here we adopt the Regge interpretation that the descritized extrinsic curvature $K$, just like the intrinsic curvature, is distributed only at the faces of the polytope-decomposition dual to $\gamma$. According to the above discussion, the value of $K_e$ may thus be expressed in either the local gauge for the source polytope or that for the target polytope, respectively as
\begin{eqnarray}
\label{corres2}
\left(e^{\bar{\xi}_e^\mu {u}_{e}\!\bar\tau_\mu\!{u}^{-1}_{e}}\,e^{ -\bar{\eta}_e^\mu {u}_{e}\!\bar\tau_\mu\!{u}^{-1}_{e}}\, e^{(\xi_e- \eta_e) {V}_e}\right)=e^{\beta K_e} \,\,\,\text{or}\,\, \,\left(e^{-\bar{\eta}_e^\mu \tilde{u}_{e}\!\bar\tau_\mu\!\tilde{u}^{-1}_{e}}\,e^{\bar{\xi}_e^\mu \tilde{u}_{e}\!\bar\tau_\mu\!\tilde{u}^{-1}_{e}}\, e^{-(\xi_e- \eta_e) \tilde{V}_e}\right)= e^{\beta K_e}\,.\nonumber\\
\end{eqnarray}
A well-known feature of a Regge hypersurface is that the the extrinsic curvature distribution at a specific face of a constituent polytope must be a vector 1-form distribution parallel to the normal of the face. This knowledge then suggests the further correspondence of
\begin{eqnarray}
\label{corres3}
\frac{1}{\beta}(\xi_e- \eta_e) {V}_e= K^{\perp}_e \,\,\,\text{or}\,\, \,\frac{1}{\beta}(\xi_e- \eta_e) \tilde{V}_e= -K^{\perp}_e,
\end{eqnarray}
when expressed in the source frame or target frame. Finally, this leaves the remaining degrees of freedom to account for $K^{/\!/}_e$ via
\begin{eqnarray}
\label{corres4}
\frac{1}{\beta}\ln(e^{\bar{\xi}_e^\mu {u}_{e}\!\bar\tau_\mu\!{u}^{-1}_{e}}\,e^{ -\bar{\eta}_e^\mu {u}_{e}\!\bar\tau_\mu\!{u}^{-1}_{e}})= K^{/\!/}_e \,\,\,\text{or}\,\, \,\frac{1}{\beta}\ln(e^{-\bar{\eta}_e^\mu \tilde{u}_{e}\!\bar\tau_\mu\!\tilde{u}^{-1}_{e}}\,e^{\bar{\xi}_e^\mu \tilde{u}_{e}\!\bar\tau_\mu\!\tilde{u}^{-1}_{e}})= K^{/\!/}_e.
\end{eqnarray}
In general, the data in $(N_e, V_e, \tilde{V}_e)$ contains information about both intrinsic and extrinsic geometry. Out of these $4D-3$ degrees of freedom of $(N_e, V_e, \tilde{V}_e)$, only $2D-1$ of them would be interpretable as intrinsic-geometry property of the D-dimensional slice, while the other $2D-2$ of them carry information about the extrinsic geometry. The extra angle $\xi_e$ is the missing ingredient necessary in order to reconstruct the $(2D-1)_{\textrm{th}}$ component of $K^{\perp}_e$. As we will demonstrate in more details, the angles $\bar{u}_e$ containing the information about the components $K^{/\!/}_e$ of $K_e$ are purely redundant variables, in views of both the Regge hypersurface geometry and gauge reduction involving the discretized simplicity constraints.

The set of angle-bivector variables $(N_e, V_e, \tilde{V}_e,\xi_e,\bar{\xi}_e^\mu)$ gives the generalization of twisted geometry parametrization for $SO(D+1)$ phase space. We will now carry out an analysis of the canonical correspondence between these variables and the LQG phase space, before coming back to provide more support on the discrete hypersurface interpretation and drawing insights on the proper treatment of the gauge reduction with the anomalous discretized simplicity constraints.

\section{Sympletic analysis of edge-simplicity constraint surface in $SO(D+1)$ LQG phase space}

\subsection{Sympletic structure of $SO(D+1)$ LQG phase space}

Recall that the phase space of $SO(D+1)$ loop quantum gravity associated with each edge of a given graph can be given by the group tangent space $TSO(D+1)$. Since this space is bundle-isomorphic to $T^*SO(D+1)$, as a phase space it enjoys the natural symplectic structure of the $SO(D+1)$ cotangent bundle. Explicitly, the bundle isomorphism $TSO(D+1) \to T^*SO(D+1) $ is given by the trivialization $( so(D+1), SO(D+1)) \to TSO(D+1)$ using a basis of right-invariant $so(D+1)$ vector fields, followed by the identification $so(D+1) \to so^*(D+1)$ that leads to the trivialization of the cotangent bundle $(so^*(D+1), SO(D+1)) \to T^*SO(D+1)$.

A right-invariant vector field $\hat X$ associated to the Lie algebra element $X\in \mathfrak{g}$, acts on a function on the group manifold via the right derivative $\nabla_X^R$ as
\begin{equation}
\nabla_X^Rf(g)\equiv \frac{d}{dt}f(e^{-tX}g)|_{t=0};
\end{equation}
under the adjoint transformation $X\mapsto -gXg^{-1}$, we obtain the corresponding left derivative
\begin{equation}
\nabla_X^Lf(g)\equiv \frac{d}{dt}f(ge^{tX})|_{t=0}=-\nabla^R_{gXg^{-1}}f(g).
\end{equation}
It is straight forward to show that the map from the right invariant vector fields $\hat{X}$ to the corresponding elements $X$ of the algebra is provided by the algebra-valued, right-invariant 1-form $dgg^{-1}$ satisfying
\begin{equation}
i_{\hat{X}}(dgg^{-1})=(\mathcal{L}_{\hat{X}}g)g^{-1}=-X,
\end{equation}
where $i$ denotes the interior product, and $\mathcal{L}_{\hat{Y}}\equiv i_{\hat{Y}}d+di_{\hat{Y}}$ denotes the Lie derivative. It is clear from the above, that a basis for $\mathfrak{g}$ is then associated to a set of right-invariant vector fields, which serves as a global tangent-space basis providing the trivialization as $(so(D+1), SO(D+1)) \to TSO(D+1)$. Moreover, there is also a (local) coordinate system $G_{IJ}: SO(D+1) \to so(D+1)$ for the base manifold $SO(D+1)$, such that for any element $g$, we have $dG_{IJ}|_g\equiv (dgg^{-1})_{IJ}$. Using such a local coordinate system, the trivialization thus locally takes the form of $(X_{IJ}, G_{IJ}) \to TSO(D+1)$. Next, to describe the cotangent fiber bundle, we note that for every $X\in\mathfrak{g}$ there is a corresponding element $h_X$ in the dual algebra $\mathfrak{g}^\ast$, which as a linear function of $Y\in\mathfrak{g}$ is defined by $$h_X(Y)\equiv \textrm{Tr}(XY ) \equiv-2\textrm{tr}(X_{IJ}\tau^{IJ}Y_{KL}\tau^{KL})=2X_{IJ}Y_{KL}\delta^{K[I}\delta^{J]L}=2X^{KL}Y_{KL},$$ with the duality map given by the non-degenerate operator $$\frac{1}{2}\textrm{Tr}(\tau^{IJ}\tau^{KL})=\delta^{K[I}\delta^{J]L}.$$ Using this ad-invariant pairing we have identified $so(D+1)$ to $so(D+1)^\ast$ and specified the trivial cotangent bundle structure $(so^*(D+1), SO(D+1)) \to T^*SO(D+1)$. Thereby, the above (local) coordinate system describes the trivialization of the cotangent bundle in the explicit form $(X^{IJ}, G_{IJ}) \to T^*SO(D+1)$.

Recognizing that by construction $(X^{IJ}, G_{IJ})$ (locally) forms the dual coordinate pair of the cotangent bundle, we can now simply read off the natural symplectic potential for $TSO(D+1)$ as
\begin{eqnarray}\label{sympotential}
\Theta\equiv X^{IJ}dG_{IJ} =\frac{1}{2}\textrm{Tr}(Xdgg^{-1}).
\end{eqnarray}
The symplectic 2-form then follows as
\begin{equation}
\Omega\equiv -d\Theta=- \frac{1}{2}d\textrm{Tr}(Xdgg^{-1})=\frac{1}{4}\textrm{Tr}(d\tilde{X}\wedge g^{-1}dg-dX\wedge dgg^{-1})
\end{equation}
where we have introduced $\tilde{X}\equiv-g^{-1}Xg$. Among the interesting phase space functions in $TSO(D+1)$ (or equivalently in $T^*SO(D+1)$), we will specifically study the ones of the form $f\equiv f(g)$ and $h_Y\equiv h_Y(X)$. From the symplectic 2-form we can compute the following important Poisson brackets among them:
\begin{equation}\label{Poiss}
\{h_Y,h_Z\}=2h_{[Y,Z]},\quad \{h_Y,f(g)\}=2\nabla^R_Yf(g),\quad \{f(g),h(g)\}=0.
\end{equation}
\textbf{Proof.} Let us identify $so(D+1)$ with $\mathbb{R}^{\frac{D(D+1)}{2}}$ via $X^i=\textrm{Tr}(\tau^iX)=h_{\tau^i}(X)$, where $i\in\{1,...,\frac{D(D+1)}{2}\}$ and $\tau^i$ is an element of the orthogonal basis of $so(D+1)$. Consider the following vector field on $T^\ast\!SO(D+1)$,
\begin{equation}
\hat{Y}\equiv\nabla^R_Y+[X,Y]^i\frac{\partial}{\partial X^i}.
\end{equation}
This vector field is such that
\begin{eqnarray}
i_{\hat{Y}}\Theta&=&- \frac{1}{2}\textrm{Tr}(XY),\\\nonumber
\mathcal{L}_{\hat{Y}}\Theta&=& \frac{1}{2}\textrm{Tr}([X,Y]dgg^{-1})- \frac{1}{2}\textrm{Tr}(X[Y,dgg^{-1}])=0.
\end{eqnarray}
Therefore we have
\begin{equation}
i_{\hat{Y}}\Omega=di_{\hat{Y}}\Theta-\mathcal{L}_{\hat{Y}}\Theta=- \frac{1}{2}d\textrm{Tr}(XY),
\end{equation}
which implies that $\hat{Y}$ is the Hamiltonian vector field of $ \frac{1}{2}h_Y(X)$ and
\begin{equation}
\{ \frac{1}{2}h_Y, \frac{1}{2}h_Z\}=\Omega(\hat{Y},\hat{Z})=- \frac{1}{2}i_{\hat{Z}}dh_Y= \frac{1}{2}h_{[Y,Z]}.
\end{equation}
Next, the Hamiltonian vector field of a function $f(g)$ on the group is
\begin{equation}
\hat{f}=- 2\nabla^R_{X^i}f\frac{\partial}{\partial X^i},
\end{equation}
since
\begin{equation}
i_{\hat{f}}\Omega= \nabla^R_{X^i}f\textrm{Tr}(\tau^idgg^{-1})\equiv -df.
\end{equation}
It is then easy to see that any two functions of the forms $f(g)$ and $h(g)$ would have a vanishing Poisson bracket as given by $\Omega_{T^\ast\!G}(\hat{X}_f,\hat{X}_h)=0$. Finally, we have
\begin{equation}
\{ \frac{1}{2}h_Y,f\}=i_{\hat{Y}}df=-i_{\hat{X}_f}dh_Y=\nabla^R_Yf.
\end{equation}
$\square$\\
We see from the brackets (\ref{Poiss}) that the Poisson action of $h_Y(X)$ generates left derivatives. Similarly, the right derivative $\{\tilde{h}_Y,f(g)\}=2\nabla^L_Yf(g)$ is generated by the action of $\tilde{h}_Y(X)\equiv \textrm{Tr}(Y\tilde{X})$ with $\tilde{X}=-g^{-1}Xg$. Finally, the two Hamiltonians commute as given by $\{h_Y,\tilde{h}_Z\}=0$.

Using the obtained Poisson brackets, one may evaluate the algebra amongst the discretized Gauss constraints, edge-simplicity constraints and vertex-simplicity constraints defined in \eqref{gaussconstr} and \eqref{simpconstr}. It turns out that $G_v\approx0$ and $S_e\approx0$ form a first class constraint system, with the algebra
\begin{eqnarray}
\label{firstclassalgb}
\{S_e, S_e\}\propto S_e\,,\,\, \{S_e, S_v\}\propto S_e,\,\,\{G_v, G_v\}\propto G_v,\,\,\{G_v, S_e\}\propto S_e,\,\,\{G_v, S_v\}\propto S_v, \quad b(e)=v,
\end{eqnarray}
where the brackets within $G_v\approx0$ is just the $so(D+1)$ algebra, and the ones within $S_e\approx0$ weakly vanish. The algebra involving the vertex-simplicity constraint are the problematic ones, with the open anomalous brackets
\begin{eqnarray}
\label{anomalousalgb}
\{S_{v,e,e'},S_{v,e,e''}\}\propto \emph{anomaly term}
\end{eqnarray}
where the terms $ \emph{anomaly term}$ are not proportional to any of the existing constraints in the phase space.

\subsection{ Symplectomorphism between edge-simplicity constraint surface and angle-bivector space}

Having discussed the symplectic structure of the $T^*SO(D+1)$ phase space, we recall the angle-bivector parametrization for the edge-simplicity constraint surface this space using the twisted-geometry variables $(V,\tilde{V},\xi, N,\bar{\xi}^\mu)\in P:=Q_{D-1}\times Q_{D-1}\times T^*S\times SO(D-1)$, where $e^{\bar{\xi}^\mu\bar{\tau}_\mu}:=\bar{u}$, and $\bar{\tau}_\mu\in so(D-1)$, $\mu\in\{1,...,\frac{(D-1)(D-2)}{2}\}$. To capture the intrinsic curvature, we have specified one pair of the $so(D+1)$ valued Hopf sections-- $u(V)$ and $\tilde{u}( \tilde{V})$-- for each edge. With the specified $u(V)$ and $\tilde{u}( \tilde{V})$, the parametrization associated with each edge is given by the map
\begin{eqnarray}\label{para}
(V,\tilde{V},\xi,N,\bar{\xi}^\mu)\mapsto(X,g):&& X=N\,V=N\,u(V)\tau_ou(V)^{-1}\\\nonumber
&&g=u(V)\,e^{\bar{\xi}^\mu\bar{\tau}_\mu}e^{\xi\tau_o}\,\tilde{u}(\tilde{V})^{-1}
\end{eqnarray}
which implies that $\tilde{X}\equiv-g^{-1}Xg=N\tilde{V}$. We first note that the map is a two-to-one double covering of the image that takes the bi-vector form $X=Nu\tau_ou^{-1}$ solving the edge-simplicity constraint $X^{[IJ}X^{KL]}=0$. Let us denote this bi-vector subset as $so(D+1)_s$, and denote the image as $T_{s}^\ast \!SO(D+1)\equiv T^\ast \!SO(D+1)|_{X^{[IJ}X^{KL]}=0}$ that is the edge-simplicity constraint surface in the phase space. Clearly, under the map introduced above from $P$ to $T_{s}^\ast \!SO(D+1)$, the two points $(V,\tilde{V},\xi, N,\bar{\xi}^\mu)$ and $(-V,-\tilde{V},-\xi,-N,\dot{\xi}^\mu)$ related with $e^{\dot{\xi}^\mu\bar{\tau}_{\mu}}=e^{-2\pi\tau_{13}}e^{\bar{\xi}^\mu\bar{\tau}_{\mu}}e^{2\pi\tau_{13}}$ and $\tau_{13}=\delta_1^{[I}\delta_3^{J]}$ are mapped to the same point $(X, g)\in T_{s}^\ast \!SO(D+1) $. A bijection map can thus be established in the region $|X|\neq0$ by selecting either branch among the two signs, leading to the corresponding one of the two inverse maps from the region with $|X|\neq0$ given by:
\begin{equation}
N=|X|,\quad V=\frac{X}{|X|}, \quad \tilde{V}=-\frac{g^{-1}Xg}{|X|},\quad \xi=\textrm{Tr}(\tau_o\ln(u^{-1}g\tilde{u})),\quad \bar{\xi}^\mu=\textrm{Tr}(\bar{\tau}^\mu\ln(e^{-\xi\tau_o}u^{-1}g\tilde{u}))
\end{equation}
or
\begin{equation}
N=-|X|,\quad V=-\frac{X}{|X|}, \quad \tilde{V}=\frac{g^{-1}Xg}{|X|},\quad \xi=-\textrm{Tr}(\tau_o\ln(u^{-1}g\tilde{u})),\quad \dot{\xi}^\mu=\textrm{Tr}(\bar{\tau}^\mu\ln(e^{\xi\tau_o}e^{-2\pi\tau_{13}}u^{-1}g\tilde{u}e^{2\pi\tau_{13}})).
\end{equation}
Thus, we have an isomorphism between the two sets
\begin{equation}\label{PSO}
P^\ast/\mathbb{Z}^2\rightarrow T_{s}^\ast \!SO(D+1)\!\setminus\!\{|X|=0\},
\end{equation}
where $P^\ast\equiv P|_{N\neq 0}$ denotes the region with ${N\neq 0}$, and the identifying $\mathbb{Z}^2$ operation is defined by
\begin{equation}
(V,\tilde{V},\xi, N,\bar{\xi}^\mu) \rightarrow (-V,-\tilde{V},-\xi,-N,\dot{\xi}^\mu)
\end{equation}
in the region $N\neq0$.

Since $P^\ast\equiv P|_{N\neq 0}$ provides a double-covering coordinate system for $ T_{s}^\ast \!SO(D+1)\!\setminus\!\{|X|=0\}$, we may use the bivector-angle variables to express the induced presymplectic structure of $T_{s}^\ast \!SO(D+1)\!\setminus\!\{|X|=0\}$ inherited from the phase space $T^*\!SO(D+1)$. First, the induced presymplectic potential can be expressed as
\begin{eqnarray}
\Theta_{T_s^\ast\!SO(D+1)} |_{|X|>0}&=& \frac{1}{2}\textrm{Tr}(Xdgg^{-1})|_{T_{s}^\ast\!SO(D+1); |X|>0}\\\nonumber
&=& \frac{1}{2}N\textrm{Tr}(u\tau_{o}u^{-1} (duu^{-1}+u(d\xi\tau_o+d\bar{\xi}^\mu\bar{\tau}_\mu)u^{-1}-ue^{\bar{\xi}^\mu\bar{\tau}_\mu}e^{\xi\tau_o} \tilde{u}^{-1}d\tilde{u}\tilde{u}^{-1}\tilde{u}e^{-\bar{\xi}^\mu\bar{\tau}_\mu}e^{-\xi\tau_o} u^{-1})) \\\nonumber
&=& \frac{1}{2}N\textrm{Tr}(Vduu^{-1})+ \frac{1}{2}Nd\xi- \frac{1}{2}N\textrm{Tr}(\tilde{V}d\tilde{u}\tilde{u}^{-1}).
\end{eqnarray}
From the point of view of the space $P$, we may extend this potential in the limit $N\to0$ and simply define
\begin{equation}
\Theta_{P}\equiv \frac{1}{2}N\textrm{Tr}(Vduu^{-1})+ \frac{1}{2}Nd\xi- \frac{1}{2}N\textrm{Tr}(\tilde{V}d\tilde{u}\tilde{u}^{-1})
\end{equation}
as the presymplectic potential in $P$. This potential gives the presympletic form $\Omega_P$ as
\begin{eqnarray}
\Omega_P=-d\Theta_P &=& \frac{1}{2}N\textrm{Tr}(Vduu^{-1}\wedge duu^{-1})-\frac{1}{2}N\textrm{Tr}(\tilde{V}d\tilde{u}\tilde{u}^{-1}\wedge d\tilde{u}\tilde{u}^{-1}) \\\nonumber
&& -\frac{1}{2}dN\wedge (d\xi+\textrm{Tr}(Vduu^{-1})-\textrm{Tr}(\tilde{V}d\tilde{u}\tilde{u}^{-1})).
\end{eqnarray}
It is clear that the $N=0$ region of the above presymplectic structure is degenerate, as expected due to the degeneracy in the parametrization itself in the ${N= 0}$ region of $T_{s}^\ast \!SO(D+1)$. More importantly, as we shall demonstrate in the next section, the induced presymplectic structure for $P^\ast$ coincides with the natural symplectic structures of the two constituent spaces --- the $Q_{D-1}$ and $T^*S^1$, while leaving the third component $SO(D-1)$ completely degenerate. Therefore, this $SO(D-1)$ component faithfully parametrizes the symplectic degeneracies of $T_{s}^\ast \!SO(D+1)\!\setminus\!\{|X|=0\}$ as a presymplectic manifold. Since the edge-simplicity constraints form a first class system with the discretized Gauss constraints, we expect the $SO(D-1)$ degenerate degrees of freedom to be generated by the first-class constraints. Indeed, we note that the induced symplectic form on $T_s^\ast\!SO(D+1)\!\setminus\!\{|X|=0\}$ given by $\Omega_{T^\ast\!SO(D+1)}\equiv -d\Theta_{T^\ast\!SO(D+1)}$ is different from $\Omega_{P^\ast}:=-d\Theta_{P^\ast}$ obtained from the induced symplectic potential $\Theta_{P^\ast}$, since the Hamiltonian vector fields of any function on $P^\ast$ given by the two symplectic forms always differ by a transformation induced by the edge-simplicity constraints. More explicitly, we can evaluate the transformations induced by the edge-simplicity constraints in the LQG discrete phase space, and obtain
\begin{equation}
\{S_e^{IJKL}, X_e\}|_{S_e=0}=0\,\,\text{and}\,\,\,\{S_e^{IJKL}, g_e\}|_{S_e=0}\propto X_e^{[IJ}(\tau^{KL]}g_e)|_{S_e=0}\propto V_e^{[IJ}(\tau^{KL]}g_e)|_{V_e=u_e \tau_0 u_e^{-1}}.
\end{equation}
Now it is easy to see that the edge-simplicity constraint transforms the holonomy $g_e$ by the left action of an $SO(D-1)$ element preserving the two vectors forming $V_e$. Then, via the parametrization \eqref{para} of $g_e$, we conclude that the edge-simplicity constraints generate the transformation of the $SO(D-1)$ angles, which are precisely the degenerate component with respected to the presymplectic form $\Omega_{P^\ast}$. Lastly, let us view the above transformations induced by $S_e^{IJKL}$ under the discrete Regge geometry interpretation proposed in section 3.2. Since the edge-simplicity constraints commute (on-shell) with the flux variables, it is clear that the transformations act trivially on the intrinsic geometry as desired. Moreover, the above shows that the transformations change only the $SO(D-1)$ angles $\bar{\xi}_e^\mu$ among the twisted-geometry variables, then according to our interpretation \eqref{corres4} the transformations act only upon $K^{/\!/}_e$, which are indeed the pure gauge components in the original Ashtekar formulation.

To go further and study the gauge reductions in the new geometric point of view, we need to compute the Poisson brackets between the twisted-geometry variables using the presymplectic form $\Omega_P$. In order to do that, in the following section we will study the Hopf sections $u(V)$ and $\tilde{u}(\tilde{V})$ in the perspectives of their contributions to the Hamiltonian fields on $P$ defined by $\Omega_P$ .

\subsection{Hopf map and Geometric action on the Hopf section}

The Hopf map is defined as a special projection map $\pi: SO(D+1)\mapsto Q_{D-1}$ with $Q_{D-1}:=SO(D+1)/(SO(2)\times SO(D-1))$, such that every element in $Q_{D-1}$ comes from the maximal subgroup of $SO(D+1)$ that fixed $\tau_{o}$. The maximal subgroup takes the form $SO(2)\times SO(D-1)$, and in the definition representation of $SO(D+1)$ the Hopf map reads
\begin{eqnarray}
\pi: \quad SO(D+1) &\rightarrow& Q_{D-1} \\\nonumber
g &\rightarrow& V(g)=g\tau_og^{-1}.
\end{eqnarray}
Note that the vector $V(g)$ is invariant under $g\mapsto g^{\alpha,\beta^\mu}=ge^{\alpha\tau_o+\beta^\mu\bar{\tau}_\mu}$, thus it is a function of $2D-2$ variables only. This result shows that $SO(D+1)$ can be seen as a bundle (we would call it the Hopf bundle) over $Q_{D-1}$ with a $SO(2)\times SO(D-1)$
fiber. On this bundle we can introduce the Hopf sections, each as an inverse map to the above projection
\begin{eqnarray}
u:\quad Q_{D-1} &\rightarrow& SO(D+1)\\\nonumber
V&\mapsto& u(V),
\end{eqnarray}
such that $\pi(u(V))=V$. This section assigns a specific $SO(D+1)$ element $u$ to each member of the $Q_{D-1}$, and it is easy to see that any given section $u$ is related to all other sections via $u^{\alpha,\alpha'^{\mu}}\equiv ue^{\alpha\tau_o+\alpha'^{\mu}\bar{\tau}_\mu}$; therefore the free angles $\{\alpha,\alpha'^{\mu}\}$ parametrize the set of all possible Hopf sections.

Let us identify $so(D+1)$ with $\mathbb{R}^{\frac{D(D+1)}{2}}$ via the representation $X=X^{IJ}$. Then, an element $V\in Q_{D-1}$ is identified with a unit bi-vector in $\mathbb{R}^{\frac{D(D+1)}{2}}$, and we have a natural action of rotations by the group $SO(D+1)$ in this space. Since this action is given via the co-adjoint representation, we can further associate each algebra element $X\in so(D+1)$ to a vector field $\hat{X}$ on $Q_{D-1}$, which acts on a function of $Q_{D-1}$ as
\begin{equation}
\mathcal{L}_{\hat{X}}f(V):=\frac{d}{dt}f(e^{-tX}Ve^{tX})|_{t=0}.
\end{equation}
Specifically in the case of linear functions we have
\begin{equation}\label{LXV}
\mathcal{L}_{\hat{X}}V=-[X,V].
\end{equation}
Next, we observe that the $SO(D+1)$ action on $Q_{D-1}$ as a symplectic manifold is Hamiltonian; by explicit calculation one can verify that $\hat{X}$ is a Hamiltonian vector field associated to the function $ \frac{1}{2}h_X(V)\equiv NV^{IJ}X_{IJ}$ on $Q_{D-1}$, and the action above can be obtained from the Poisson bracket between $V$ and $ \frac{1}{2}h_X$, which results to
\begin{equation}\label{Haac}
\{ \frac{1}{2}h_X,V\}=N\Omega_{Q}(\hat{X},\hat{V})=-[X,V]=\mathcal{L}_{\hat{X}}V.
\end{equation}

We are especially interested in the action of the algebra on the Hopf section. Let us
first note that
\begin{equation}
\label{LX}
\mathcal{L}_{\hat{X}}V(u)=(\mathcal{L}_{\hat{X}}u)\tau_ou^{-1} +u\tau_o(\mathcal{L}_{\hat{X}}u^{-1})=[(\mathcal{L}_{\hat{X}}u)u^{-1}, V(u)].
\end{equation}
Comparing this with (\ref{LXV}), we deduce that
\begin{equation}\label{LXuu}
(\mathcal{L}_{\hat{X}}u)u^{-1}=-X+V(u)F_X(V)+\sum_{\mu}\bar{V}_\mu(u)L^\mu_X(V),
\end{equation}
where $\bar{V}_\mu(u)\equiv u\bar{\tau}_\mu u^{-1}$, $F_X(V)$ and $L^\mu_X(V)$ are functions of $V\in Q_{D-1}$, with both $V(u)F_X(V)$ and $\bar{V}_\mu(u)L^\mu_X(V)$ commuting with the element $V(u)$ for all $\mu$.\\ \\
\textbf{Lemma.}
Define $\bar{V}^\mu_{IJ}=u(V)\bar{\tau}_{IJ}^\mu u(V)^{-1}$, the solution function $L^{IJ}\equiv L: Q_{D-1}\mapsto so(D+1)$ of the equations
\begin{equation}
\textrm{Tr}(Lduu^{-1})=0, \quad L^{IJ}V_{IJ}=1,\quad L^{IJ}\bar{V}^\mu_{IJ}=0,\ \forall \mu,
\end{equation}
appears in the Lie derivative of the Hopf map section $u(V)$ as,
\begin{equation}
L_X=2F_X
\end{equation}
and it satisfies the key coherence identity
\begin{equation}\label{LXY}
\mathcal{L}_{\hat{X}}L_Y-\mathcal{L}_{\hat{Y}}L_X=L_{[X,Y]}.
\end{equation}
Finally, the general solution to this identity satisfying the conditions $L^{IJ}V_{IJ}= 1, L^{IJ}\bar{V}^\mu_{IJ}=0$ is given by
\begin{equation}\label{gf}
L'=L+d\alpha
\end{equation}
where $\alpha$ is a function on $Q_{D-1}$.

\textbf{Proof.}

Takeing the interior product of an arbitrary vector field $\hat{X}$ with the defining expression $\textrm{Tr}(Lduu^{-1})=0$ and recalling that by definition of Lie derivative $(\mathcal{L}_{\hat{X}}u)u^{-1}=i_{\hat{X}}(duu^{-1})$, we have
\begin{equation}
0=i_{\hat{X}}\textrm{Tr}(Lduu^{-1})=\textrm{Tr}(L(\mathcal{L}_{\hat{X}}u)u^{-1}) =-\textrm{Tr}(LX)+F_X\textrm{Tr}(LV)=-L_X+2F_X,
\end{equation}
where we used $L^{IJ}V_{IJ}=1, L^{IJ}\bar{V}^\mu_{IJ}=0,$ and (\ref{LXuu}). Hence we proved $2F_X=L_X$.

To prove (\ref{LXY}) we first observe that
\begin{eqnarray}
\mathcal{L}_{\hat{X}}(duu^{-1})&=&i_{\hat{X}}(duu^{-1}\wedge duu^{-1})+d[(\mathcal{L}_{\hat{X}}u)u^{-1}]\\\nonumber
&=&[-X+\frac{1}{2}VL_X+\sum_{\mu}\bar{V}_\mu L^\mu_X,duu^{-1}]+d(-X+\frac{1}{2}VL_X+\sum_{\mu}\bar{V}_\mu L^\mu_X)\\\nonumber
&=&\frac{1}{2}VdL_X+\bar{V}_\mu dL^\mu_X-[X,duu^{-1}],
\end{eqnarray}
where we used the definition of Lie derivative in the first equality, (\ref{LXuu}) in the second and $dV=[duu^{-1},V]$ with $d\bar{V}^\mu=[duu^{-1},\bar{V}^\mu]$ in the third. The above then leads to
\begin{equation}
0=\mathcal{L}_{\hat{X}}\textrm{Tr}(Lduu^{-1})=\textrm{Tr}((\mathcal{L}_{\hat{X}}L-[L,X])duu^{-1}) +dL_X
\end{equation}
with the help of the equalities $L^{IJ}V_{IJ}=1$ and $L^{IJ}\bar{V}^\mu_{IJ}=0$.

Finally, by taking the interior product of the last equation with $\hat{Y}$ we get
\begin{eqnarray}
\mathcal{L}_{\hat{Y}}L_X&=& \textrm{Tr}((\mathcal{L}_{\hat{X}}L-[L,X] )(Y-\frac{1}{2}VL_Y-\sum_{\mu}\bar{V}_\mu L^\mu_Y))\\\nonumber
&=&\mathcal{L}_{\hat{X}}L_Y-L_{[X,Y]}-\frac{1}{2}L_Y(\textrm{Tr}((\mathcal{L}_{\hat{X}}L)V) -\textrm{Tr}(L[X,V]))-\sum_{\mu} L^\mu_Y(\textrm{Tr}(\mathcal{L}_{\hat{X}}L\bar{V}_\mu) -\textrm{Tr}(L[X,\bar{V}_\mu]))\\\nonumber
&=&\mathcal{L}_{\hat{X}}L_Y-L_{[X,Y]}-\frac{1}{2}L_Y\mathcal{L}_{\hat{X}}(\textrm{Tr}(LV) )-\sum_{\mu} L^\mu_Y\mathcal{L}_{\hat{X}}\textrm{Tr}(L\bar{V}_\mu)
\end{eqnarray}
and since the last two terms vanish, we obtain the coherence identity (\ref{LXY}).

Suppose we have another solution $L'$ to the coherence identity and also the conditions $L^{IJ}V_{IJ}=1$ and $L^{IJ}\bar{V}^\mu_{IJ}=0$. Using the 1-form $\beta\equiv -\textrm{Tr}(L'duu^{-1})$ we see can that its contraction with $\hat{X}$
\begin{equation}
\label{Ltransf}
\beta_X\equiv i_{\hat{X}}\beta=-\textrm{Tr}(L'(\mathcal{L}_{\hat{X}}u)u^{-1})=L'_X-L_X
\end{equation}
is the difference between the two solutions and thus also a solution to the coherence identity. This, together with the
definition of the differential $i_{\hat{X}}i_{\hat{Y}}d\beta=\mathcal{L}_{\hat{Y}}\beta_X -\mathcal{L}_{\hat{X}}\beta_Y+\beta_{[X,Y]}$, implies that $d\beta=0$, which means that there exist a function $\alpha$ locally such that $\beta=d\alpha$ at least, and thus $L'_X=L_X+\mathcal{L}_{\hat{X}}\alpha$. This proves the gauge freedom (\ref{gf}).

$\square$

Finally, let us recall that the freedom in choosing the Hopf section lies in the two function parameters $\alpha(V)$ and $\alpha'^{\mu}(V)$ in the expression $u'(V)\equiv u(V)e^{\alpha(V)\tau_o+\alpha'^{\mu}(V)\tau_\mu}$ for all possible choices of the sections. Applying the equation \eqref{LXuu} to this $u'$, we immediately get $L'_X= L_X+ i_{\hat{X}}d\alpha$. Referring to \eqref{Ltransf}, we see now that the set of functions $L$ satisfying the above three key conditions is exactly the set of the function coefficients for the component of $(du)u^{-1}$ in the $V$ direction, given under all possible choices of the Hopf section $u$. Applying these conditions in the presympletic form $\Omega_P$, we will now identify the Hamiltonian fields in $P$ and compute the Poisson brackets.

\subsection{Computation of Hamiltonian vector fields in pre-symplectic manifold $P$}

Recall that we have obtained the pre-symplectic potential $\Theta_{P}:= \frac{1}{2}N\textrm{Tr}(Vduu^{-1})+ \frac{1}{2}Nd\xi- \frac{1}{2}N\textrm{Tr}(\tilde{V}d\tilde{u}\tilde{u}^{-1})$ induced from the edge-simplicity constraint surface in the $SO(D+1)$ phase space. The potential defines a presympletic form $\Omega_P$ as
\begin{eqnarray}
\Omega_P=-d\Theta_P &=& \frac{1}{2}N\textrm{Tr}(Vduu^{-1}\wedge duu^{-1})-\frac{1}{2}N\textrm{Tr}(\tilde{V}d\tilde{u}\tilde{u}^{-1}\wedge d\tilde{u}\tilde{u}^{-1}) \\\nonumber
&& -\frac{1}{2}dN\wedge (d\xi+\textrm{Tr}(Vduu^{-1})-\textrm{Tr}(\tilde{V}d\tilde{u}\tilde{u}^{-1})).
\end{eqnarray}
To compute the associated Poisson brackets, we first need to compute the Hamiltonian vector fields on $P$. Let us denote the Hamiltonian vector field for the function $f$ as $\chi_f$ , where $f\in \{N, \xi, h_X\equiv NV_X, \tilde{h}_X\equiv N\tilde{V}_X\}$. Using the definition and $i_{\chi_f}\Omega_P=-df$, in the $N\neq0$ region the vector fields could be checked to be given by
\begin{eqnarray}\label{vf}
\chi_{h_X} &=& 2\hat{X}-L_X(V)\partial_\xi,\quad \chi_{\tilde{h}_X} = - 2 \hat{\tilde{X}}-L_X(\tilde{V})\partial_\xi, \\\nonumber
\chi_N&=& -2\partial_\xi,\quad \quad \quad \quad \quad \chi_\xi=2\partial_N+\frac{4}{N}\widehat{[L,V]}+\frac{4}{N}\widehat{[L,\tilde{V}]}.
\end{eqnarray}
Here $\hat{X}$ and $\widehat{[L,V]}$ are the vector fields generating the adjoint action on $Q_{D-1}$ labelled by $V$, associated respectively to the algebra elements $X$ and $[L(V),V]$. Similarly, $\hat{\tilde{X}}$ and $\widehat{[L,\tilde{V}]}$ are the vector fields generating the adjoint action on $Q_{D-1}$ labelled by $\tilde{V}$, associated respectively to the algebra elements $X$ and $[L(\tilde{V}),\tilde{V}]$.\\ \\
\textbf{Proof.} To check the first equation of (\ref{vf}), we first note that for a constant $X$ we have
\begin{equation}
i_{\hat{X}}\Omega_P=-\frac{1}{2}\textrm{Tr}(d(NV)X)+\frac{1}{4}L_X(V)dN.
\end{equation}
Since we have $i_{\partial_\xi}\Omega_P=\frac{1}{2}dN$, the first equation of (\ref{vf}) follows immediately. The computation for $\chi_{\tilde{h}_X}$ is similar with an opposite sign due to the reversal of the orientation. To check for $\chi_\xi$, we first evaluate
\begin{equation}
i_{\partial_N}\Omega_P=-\frac{1}{2}d\xi-\frac{1}{2}\textrm{Tr}(Vduu^{-1})+\frac{1}{2}\textrm{Tr}(\tilde{V}d\tilde{u}\tilde{u}^{-1}),
\end{equation}
and then we have
\begin{eqnarray}
i_{\widehat{[L,V]}}\Omega &=& -\frac{1}{2}N \textrm{Tr}([V,[L,V]]duu^{-1}) -\frac{1}{2}dN\textrm{Tr}((V-L)[L,V])\\\nonumber
&=& - \frac{1}{2}N \textrm{Tr}(\frac{1}{4}(L-\textrm{Tr}(LV)V)duu^{-1}) -\frac{1}{2}dN\textrm{Tr}((V-L)[L,V])= \frac{1}{4}N \textrm{Tr}(Vduu^{-1}),
\end{eqnarray}
where we decomposed $L$ as $L=(L-\textrm{Tr}(LV)V)+\textrm{Tr}(LV)V$ and used the definitional properties of $L$. A similar calculation shows that
\begin{eqnarray}
i_{\widehat{[L,\tilde{V}]}}\Omega &=& -\frac{1}{4}N \textrm{Tr}(\tilde{V}d\tilde{u}\tilde{u}^{-1}),
\end{eqnarray}
and thus the last equation of (\ref{vf}) follows. $\square$

Let us now address the degeneracy of $\Omega_P$ resulting to the non-uniqueness of the Hamiltonian vector fields. While $\Omega_P$ is trivally closed as coming from a local symplectic potential, it has degeneracies in the directions tangent to the $SO(D-1)$ fiber and also in the boundary region with $N=0$. There are mainly two ways to reduce the manifold $P$ to obtain a sympletic manifold. The first way is to simply consider a new space $P^\ast:=P|_{N\neq0}$ and then reduce it respected to the $SO(D-1)$ fiber, then the result would be a $(4D-2)$-dimensional sympletic manifold denoted by $\check{P}^\ast$. The second way is to reduce the pre-symplectic manifold by the kernel of $\Omega_P$, i.e. to consider the quotient manifold $\bar{\check{P}}\equiv P/\textrm{Ker}(\Omega_P)$; the result would be a symplectic manifold with non-degenerate 2-form given by the quotient projection of $\Omega_P$.

In obtaining the space $\bar{\check{P}}$, we have introduced the equivalence classes under the equivalence relation $p\sim p'$ whenever $p'=e^{\hat{D}}p$, with $\hat{D}\in \textrm{Ker}(\Omega_P)$ and $p, p'\in P$. The operation is thus determined by the vector fields in the kernel of $\Omega_P$. Since it is obvious that all tangent vector fields $\hat{T}_{SO(D-1)}$ of the fiber $SO(D-1)$ belong to $\textrm{Ker}(\Omega_P)$, we may first construct $\check{P}=P/\hat{T}_{SO(D-1)}=Q_{D-1}\times T^\ast\!S^1\times Q_{D-1}$. Then, to remove the remaining kernel in the region with $N=0$, we look for the vector fields preserving the region while having the interior products with $\Omega_P$ proportional to $N$. The set of such vector fields turn out to be given by
\begin{equation}
\hat{D}_X\equiv \chi_{h_X}-\chi_{\tilde{h}_Y},
\end{equation}
where $Y=-g^{-1}Xg$ with $g=ue^{\xi\tau_0}e^{\bar{\xi}^\mu\bar{\tau}_\mu}\tilde{u}^{-1}$ being a group element rotating $V$ to $\tilde{V}=-g^{-1}Vg$. Indeed, using the fact that $V_X=\tilde{V}_Y$, the interior product of the field with the symplectic 2-form is
\begin{equation}
i_{\hat{D}_X}\Omega_P=-d(NV_X-N\tilde{V}_Y)-N\textrm{Tr}(\tilde{V}dY) =-N\textrm{Tr}([V,X]dgg^{-1}),
\end{equation}
which vanishes at $N=0$. Next, to find the equivalence class generated by the vector fields $\hat{D}_X $, we note that the actions of the fields should rotate jointly the vectors $V$ and $\tilde{V}$, that is we have $\hat{D}_X(V)=-[X,V]$, $\hat{D}_X(\tilde{V})=-g^{-1}[X,V]g$. Further, the actions preserves the group element $g$, as demonstrated by the fact that
\begin{equation}
\hat{D}_X(g)=-Xg-gY=0.
\end{equation}
Therefore, given $p\equiv (V,\tilde{V},0,\xi)$ and $p'\equiv(V',\tilde{V}',0,\xi')$, we have $ p'\sim p$ if and only if the two are related by a joint rotation in $V$ and $\tilde{V}$ and a $g$-preserving translations in $\xi$. The two copies of $Q_{D-1}$ at the ends of each edge are thus identified under this equivalence relation, and after the quotient we are left with a manifold $SO(D+1)/SO(D-1)$ parametrized by only $V$ and $\xi$.

Let us observe that the two quotient operations with respected to $\hat{T}_{SO(D-1)}$ and $\hat{D}_X$ commute, since $\hat{D}_X$ doesn't change $e^{\xi^\mu\bar{\tau}_\mu}\in SO(D-1)$ which is the degrees of freedom reduced by $\hat{T}_{SO(D-1)}$. This fact can be illustrated as

$$\xymatrix{}
\xymatrix@C=2.5cm{
P
\ar[d]^{\hat{T}_{SO(D-1)}} \ar[r]^{\hat{D}_X} & \bar{P}\ar[d]^{\hat{T}_{SO(D-1)}}\\
\check{P} \ar[r]^{\hat{D}_X}&\bar{\check{P}} }$$
where $\bar{P}$ span $P^\ast$ for $N\neq0$ and $SO(D+1)$ for $N=0$; Similarly, $\bar{\check{P}}$ span $\check{P}^\ast$ for $N\neq0$ and $SO(D+1)/SO(D-1)$ for $N=0$.

Finally, let us point out that the symplectic potential is invariant under the $\mathbb{Z}_2$ transformation
\begin{equation}\label{inverse}
(V,\tilde{V},N,\xi,\bar{\xi}^{\mu})\rightarrow (-V,-\tilde{V},-N,-\xi,\dot{\xi}^\mu).
\end{equation}
This can be seen via the transformations of the Hopf sections in the form of $u\rightarrow ue^{2\pi\tau_{13}}$ and $\tilde{u}\rightarrow \tilde{u}e^{2\pi\tau_{13}}$, with $\tau_{13}=\delta_1^{[I}\delta_3^{J]}$. Clearly these transformations leave $\Theta_P$ invariant since $d(ue^{2\pi\tau_{13}})(ue^{2\pi\tau_{13}})^{-1}=duu^{-1}$. Hence (\ref{inverse}) is a canonical transformation, and both $\check{P}^\ast/\mathbb{Z}_2$ and $\bar{\check{P}}/\mathbb{Z}_2$ are again symplectic manifolds.

\subsection{Consistency with natural Poisson structures of constituent spaces}

We have seen that the manifold $P=Q_{D-1}\times Q_{D-1}\times T^\ast\!S^1\times SO(D-1)$, viewed essentially as the edge-simplicity constraint surface of the LQG phase space, is equipped with the induced pre-symplectic potential $\Theta_{P}$. On the other hand, the space is also a product space of the components $Q_{D-1}$ and $T^\ast\!S^1$ each having a natural phase space structure. Therefore, the product space $P$ is also endowed with a class of natural Poisson structures given by the consistent gluing of the constituent spaces' symplectic structures. As it turns out, the Poisson structure given by $\Theta_{P}$ indeed belongs to such a class.

The natural phase space structure of the constituent spaces $Q_{D-1}$ and $T^\ast\!S^1$ are well-known, and they are given by:
\begin{itemize}
\item The cotangent bundle $T^\ast S^1$ with the symplectic 2-form $\Omega_{T^\ast S^1}:=\frac{1}{2}dN\wedge d\xi$, giving the Poisson bracket $\{\xi,N\}=2$.
\item The manifold $Q_{D-1}$ with the natural invariant Kahler metric and the corresponding Kahler form $\Omega_Q$, which is induce from the the standard Hermitian metric on $\mathbb{C}^{D+1}$ and re-scaled into the form $\pm\Omega_{N^2/2}:=\pm N\Omega_Q$. The sympletic form $\pm N\Omega_Q$ gives the Poisson brackets $\{NV^{IJ}, NV^{KL}\}=\pm\frac{N}{2}(\delta^{IL}V^{JK}+\delta^{ JK}V^{IL }-\delta^{IK}V^{JL}-\delta^{JL}V^{IK})$, where it becomes clear that $N=\sqrt{2NV^{IJ}NV_{IJ}}$ is a Casimir quantity satisfying $\{N,NV_{IJ}\}=0$.
\end{itemize}
Using the pre-symplectic potential $\Theta_{P}$, one could compute the Poisson brackets and obtain
\begin{eqnarray}\label{OmegaQ}
&&\{\xi,N\}=2,\nonumber\\
&&\{NV^{IJ}, NV^{KL}\}=\frac{N}{2}(\delta^{IL}V^{JK}+\delta^{ JK}V^{IL }-\delta^{IK}V^{JL}-\delta^{JL}V^{IK}),\nonumber\\
&&\{N\tilde{V}^{IJ}, N\tilde{V}^{KL}\}=-\frac{N}{2}(\delta^{IL}\tilde{V}^{JK}+\delta^{ JK}\tilde{V}^{IL }-\delta^{IK}\tilde{V}^{JL}-\delta^{JL}\tilde{V}^{IK}),\nonumber\\
&&\{V^{IJ},N\}= \{\tilde{V}^{IJ},N\}=0,
\end{eqnarray}
\begin{eqnarray}\label{brac2}
\{V^{IJ},\tilde{V}^{KL}\}=0,
\end{eqnarray}
and
\begin{equation}\label{brac3}
\{\xi^\mu,\ \cdot\ \}=0, \quad\forall \ \cdot \,.
\end{equation}
From the above, the $\Theta_{P}$ indeed endows the source and target $Q_{D-1}$ spaces respectively with the symplectic forms $N\Omega_Q$ and $-N\Omega_Q$. Also, from Eq.\eqref{brac2} the two spaces truly Poisson commute. As for the space $T^*S^1$, the induced symplectic form is also identical with $\Omega_{T^\ast S^1}$. Lastly, the vanishing brackets in Eq.\eqref{brac3} indicate the degeneracy in $\Theta_{P}$ in the $SO(D-1)$ directions. Separately in the sympletic manifolds $T^\ast S^1$ and $Q_{D-1}$, the Hamiltonian vector fields of the functions $\{\,h_X, \,\tilde{h}_X,\, N,\, \xi\,\}$ generating the above brackets can be obtained respectively according to $\Omega_{T^\ast S^1}$ and $\Omega_Q$. In comparison, the Hamiltonian vector fields in $P$ of the same functions according to $\Omega_P$ clearly differ by the terms depending on the $L$ as given in \eqref{vf}. As expected, these difference terms are generated by $\Omega_P$ via its mixing components between $T^\ast S^1$ and $Q_{D-1}$, which in turn is a result of $N$ becoming a phase space degree of freedom in $P$.

The Poisson brackets given by $\Theta_{P}$ between $\xi$ and $V$, or the ones between $\xi$ and $\tilde{V}$, turn out to be non-trivial. The results of the brackets are given by a function $L: Q_{D-1}\rightarrow so(D+1)$ in the form
\begin{equation}\label{brac4}
\{\xi,NV^{IJ}\}\equiv L^{IJ}(V), \quad \{\xi,N\tilde{V}^{IJ}\}\equiv L^{IJ}(\tilde{V}).
\end{equation}
Remarkably, the equations \eqref{brac4} taken as the definition equations for the function $L$, together with the brackets \eqref{OmegaQ}, already constrained the set of possible $L^{IJ}$ to be exactly the set of results of the brackets $\{\xi,NV^{IJ}\}$ and $ \{\xi,N\tilde{V}^{IJ}\}$ given by the potential $\Theta_{P}$ corresponding to our choice of the Hopf sections. This result can be verified by the fact that, the function $L$ defined by Eqs.\eqref{brac4} is constrained by three conditions given by the above Poisson brackets \eqref{OmegaQ}, and these three conditions are exactly the definition of $L$ in \textbf{Lemma} in section 4.3, which can be illustrated as follows. The first of the conditions comes from the equation
\begin{equation}
NV_{IJ}L^{IJ}=NV_{IJ}\{\xi,NV^{IJ}\}=\frac{1}{4}\{\xi,N^2\} =\frac{1}{2}N\{\xi,N\}=N,
\end{equation}
which gives the normalization condition $L^{IJ}(V)V_{IJ}=1$. The second condition comes from that
\begin{equation}
NV^{IJ}L^{KL}\epsilon_{IJKL\bar{M}}=NV^{IJ}\{\xi,NV^{KL}\}\epsilon_{IJKL\bar{M}} =\frac{1}{2}\{\xi,NV^{IJ}NV^{KL}\}\epsilon_{IJKL\bar{M}}=0,
\end{equation}
where we use the fact that $V$ as bi-vector satisfies $V^{IJ}V^{KL}\epsilon_{IJKL\bar{M}}=0$, with $\bar{M}$ being a $(D-3)-$tuple asymmetry index. This result implies the orthogonality condition $L^{IJ}(V)\bar{V}^\mu_{IJ}(V)=0, \forall \mu$, where $\bar{V}^\mu_{IJ}(V)\tau^{IJ}\in so(D+1)$ denotes the basis members that commutes with $V_{IJ}\tau^{IJ}\in so(D+1)$. Finally, the third constraint just comes from the Jacobi identity
\begin{equation}
\{\xi,\{NV^{IJ},NV^{KL}\}\}+\{NV^{IJ},\{NV^{KL},\xi \}\}+\{NV^{KL},\{\xi,NV^{IJ}\}\}\equiv0,
\end{equation}
from which we get the following coherence identity,
\begin{equation}\label{coh1}
\{NV^{IJ},L^{KL}(V)\}- \{NV^{KL},L^{ IJ}(V)\}\equiv \frac{1}{2}(\delta^{IL}L^{JK}(V)+\delta^{ JK}L^{IL }(V)-\delta^{IK}L^{JL}(V)-\delta^{JL}L^{IK}(V)).
\end{equation}
Similarly, the we have the conditions $L^{IJ}(\tilde{V})\tilde{V}_{IJ}=1$, $L^{IJ}(\tilde{V})\bar{V}^\mu_{IJ}(\tilde{V})=0, \forall \mu$ and
\begin{equation}\label{coh2}
\{N\tilde{V}^{IJ},L^{KL}(\tilde{V})\}- \{N\tilde{V}^{KL},L^{ IJ}(\tilde{V})\}\equiv -\frac{1}{2}(\delta^{IL}L^{JK}(\tilde{V})+\delta^{ JK}L^{IL }(\tilde{V})-\delta^{IK}L^{JL}(\tilde{V})-\delta^{JL}L^{IK}(\tilde{V})).
\end{equation}
The Hamiltonian action \eqref{Haac} can be used to write the coherence identity (\ref{coh1}) and \eqref{coh2} as an identity involving
Lie derivatives: contracting (\ref{coh1}) and \eqref{coh2} with $X^{IJ}$ and $Y^{KL}$ , we get
\begin{equation}
\mathcal{L}_{\hat{X}}L_Y-\mathcal{L}_{\hat{Y}}L_X=L_{[X,Y]},
\end{equation}
where $L_X\equiv \textrm{Tr}(LX)$ is the component of $L$ along the algebra element $X$. Now it is easy to see these three conditions makes the \textbf{Lemma} above applicable and we can verify the result given in the beginning of this paragraph.

\section{Scheme of discretized Gauss constraints and simplicity constraints reduction procedure}

So far we have discussed the phase space structure mainly associated with a single edge of the graph $\gamma$, for studying the edge-simplicity constraint surface. To carry on the constraint reduction including the Gauss constraint $G_v^{IJ}\approx0$ and vertex-simplicity constraint $S_v^{IJKL}\approx0$, we should now switch to the discrete phase space corresponding to the full graph $\gamma$. Clearly, this phase space is just given by the direct product $\mathcal{P}_\gamma\equiv\times_e T^\ast SO(D+1)_e$, with any two flux-holonomy variables associated with distinct edges Poisson commuting with each other. Then, by solving the edge-simplicity constraint equations on all of the edges of $\gamma$, the above study can be applied to the result constraint surface $\mathcal{P}^{\text{s}}_\gamma\equiv\times_e T_{\text{s}}^\ast SO(D+1)_e$ in a direct manner.

Recall that, the set $\{G_v^{IJ}\approx0, S_e^{IJKL}\approx0\}$ of the discretized Gauss and edge-simplicity constraints form a first class constraint system in $\mathcal{P}_\gamma$, with the algebra given in \eqref{firstclassalgb}. Therefore, we may perform a standard sympletic reduction with respect to this constraint system. Then, we may treat the vertex-simplicity constraint $S_v^{IJKL}\approx0$ as additional conditions, selecting from the reduced phase space the correct physical degrees of freedom. Now we proceed with the reductions upon $\mathcal{P}_\gamma$ through the following steps.

\begin{itemize}

\item{ \textbf{Symplectic reduction with respected to edge-simplicity constraint $ S_e^{IJKL}\approx0$}

From our previous analysis, the edge-simplicity constraint surface $\mathcal{P}^{\text{s}}_\gamma$ in $\mathcal{P}_\gamma$ would be given by $\times_e T_{\text{s}}^\ast SO(D+1)_e$, which is related to the fullangle-bivector space $P_\gamma$ defined as
\begin{equation}
P_\gamma\equiv \times_e P_e,\quad P_e={Q}^e_{D-1}\times Q^e_{D-1}\times T^\ast S_e^1\times SO(D-1)_e,
\end{equation}
where each $D(D+1)-\frac{(D-1)(D-2)}{2}$ dimensional space $P_e$ is described by the coordinates $(N_e,V_e,\tilde{V}_e,\xi_e,\bar{u}_e)$. Following our analysis above, we conclude that $P_\gamma|_{N_e\neq 0}$ provides a double-covering coordinatization for $\mathcal{P}^{\text{s}}_\gamma|_{X_e\neq 0}$, and the symplectomorphism (up to some gauge transformation)
 \begin{equation}
 \mathcal{P}^{\text{s}}_\gamma\cong \bar{P}_\gamma/\mathbb{Z}_2,
 \end{equation}
with $ \bar{P}_\gamma:=\times_{e\in\gamma}\bar{P}_e$ and $\bar{P}_e$ as defined in section 4.4.
Moreover, the the gauge orbits generated by edge-simplicity constraints in $\mathcal{P}_\gamma$ correspond to the degrees of freedom of $\bar{u}_e$. Therefore, the resulted reduced phase space $\mathcal{P}^{\text{S}}_\gamma$ with respected to edge-simplicity constraint can be characterized by the symplectomorphism
\begin{equation}
\label{red1}
\mathcal{P}^{\text{S}}_\gamma\cong\bar{\check{P}}_\gamma/\mathbb{Z}_2,
\end{equation}
with $\bar{\check{P}}_\gamma:=\times_{e\in \gamma}\bar{\check{P}}_e$, $\check{P}_\gamma:=\times_{e\in \gamma}\check{P}_e$ and $\check{P}_e:=Q^e_{D_1}\times Q^e_{D-1}\times T^*S_e^1$ and $\bar{\check{P}}_e$ as defined in section 4.4.
In particular, the reduced angle-bivector variables $(N_e,V_e,\tilde{V}_e,\xi_e,)$ provide a double-covering coordinatization for the reduced phase space $\mathcal{P}^{S}_\gamma$ in the ${X_e\neq 0}$ region.
 }

\item{\textbf{Symplectic reduction with respected to discretized Gauss constraints $G_v^{IJ}\approx0$}

Recall that the discretized Gauss constraints acting upon $\mathcal{P}_\gamma$ take the form $G_v=\sum_{e|s(e)=v}X_e+\sum_{e|t(e)=v}\tilde{X}_e\approx0$, and it is straight forward to see that the constraints they induce in $\mathcal{P}^{\text{S}}_\gamma$ are just the closure constraints $C_v:=\sum_{e|s(e)=v}N_eV_e+\sum_{e|t(e)=v}N_e\tilde{V}_e\approx0$ we mentioned in the beginning. The symplectic reduction inside $\mathcal{P}^{\text{S}}_\gamma$ can be perform using the closure constraint with the results given by the analysis described section $3.1$. Utilizing the solutions \eqref{gausssolution}, the obtained reduced phase space $\mathcal{P}^{\text{S,G}}_\gamma$ is characterized by the symplectomorphism
\begin{equation}
\label{red2}
\mathcal{P}^{\text{S,G}}_\gamma\cong \bar{\check{H}}_\gamma/\mathbb{Z}_2,
\end{equation}
where we define
\begin{equation}\label{quo}
\bar{\check{H}}_\gamma:=\bar{\check{P}}_\gamma/\!/SO(D+1)^{V(\gamma)},\quad\check{H}_\gamma:=\check{P}_\gamma/\!/SO(D+1)^{V(\gamma)}=\left(\times_e T^\ast S_e^1\right)\times \left(\times_v \mathfrak{P}_{\vec{N}_v}\right)
\end{equation}
 with $V(\gamma)$ being the number of the vertices in $\gamma$.
Observe that the double quotient operation in (\ref{quo}) is ``non-local" in terms of the original phase space variables, due to the fact that the variables across the two connected vertices for each edge-subspace are correlated by the condition $X_{-e}=-g^{-1}_e X_eg_e$. This technical difficulty is removed by the parametrization (\ref{para}), through which the bi-vectors $V_e$ and $\tilde{V}_e$ are assigned independently, with the relation $X_{-e}=-g^{-1}_e X_eg_e$ implicitly ensured by the definition of the angle variables. The imposition of the closure constraints and the quotient by $SO(D+1)^{V_\gamma}$ can then be taken at each of the vertices separately. Further, the reduced space carries the $T^\ast\!S^1$ degrees of freedom at every edge in the following manner. The $SO(D+1)^{V_\gamma}$ gauge orbits in this context are generated by the closure constraints acting on the remaining connection variables in $\mathcal{P}^{S}_\gamma$--- $\xi_e$; according to \eqref{brac4} the actions are given by
\begin{equation}\label{double action}
 \{\xi_e, C_{s(e)}^{IJ}\}=L^{IJ}(V_e),\quad \{\xi_e, C_{t(e)}^{IJ}\}=L^{IJ}(\tilde{V}_e).
\end{equation}
Since $g_e$ and $h^\Gamma_e$ transform identically as an $SO(D+1)$ holonomy over $e$ and notice their decomposition \eqref{hgamma}, referring to \eqref{decomp2} we infer that $\xi_e$ and $\eta_e$ behave the same under the transformations by the closure constraint:
\begin{equation}\label{double action2}
 \{\eta_e, C_{s(e)}^{IJ}\} = \{\xi_e, C_{s(e)}^{IJ}\}=L^{IJ}(V_e),\quad \{\eta_e, C_{t(e)}^{IJ}\} =\{\xi_e, C_{t(e)}^{IJ}\}=L^{IJ}(\tilde{V}_e).
\end{equation}
This implies that the extrinsic curvature 1-form $K^{\perp}_e$ identified in \eqref{corres3} is indeed $SO(D+1)$ invariant. Hence, assuming that the graph $\gamma$ is such that $\eta_e$  can be given globally without ambiguities to ensure that the Levi-Civita holonomy $h^{\Gamma}_e$ is properly defined to capture the intrinsic curvature by Eq.\eqref{hgamma}, we may use the gauge invariant $\xi^o_e:=\xi_e-\eta_e$ in place of the $\xi_e$ in $(N_e,\xi_e)\in T^\ast\! S^1$ and obtain the description of the $SO(D+1)$-invariant degrees of freedom in $T^\ast\!S^1$ under the coordinates $(N_e,\xi^o_e)$.}

\item{ \textbf{Imposing vertex-simplicity constraints $S_v^{IJKL}\approx0$}

As mentioned, here we treat the vertex-simplicity constraint $S_v^{IJKL}\approx0$ as second-class constraints in selecting the physical states of the discrete geometries from which the ADM data can be recovered.  In the space $\mathcal{P}^{\text{S,G}}_\gamma$ and $\check{H}_\gamma$, the vertex-simplicity constraints take the form $S_v^{IJKL}\equiv V^{[IJ}_{e_\imath}V^{KL]}_{e_\jmath}\approx0, $ ( $\forall e_\imath, e_\jmath: b(e_\imath)= b(e_\jmath)=v$). Denoting the subspace satisfying $S_v^{IJKL}=0$ as $\check{H}^{\textrm{s.}}_\gamma \subset \check{H}_\gamma$, we refer again to the results in Sec. $3.1$ and find that the subspace is characterized by
\begin{equation}
\label{red3}
 \check{H}^{\textrm{s.}}_\gamma=\left(\times_e T^\ast S_e^1\right)\times\left(\times_v \mathfrak{P}_{\vec{N}_v}^{\textrm{s.}}\right).
\end{equation}
To finalize our procedure, we divide $\check{H}^{\textrm{s.}}_\gamma$ by the kernel of the reduced symplectic 2-form $\Omega_\gamma\equiv \times_e\Omega_{P_e}/\!/SO(D+1)^{V_\gamma}$, to remove the artificial degeneracy resulted from the parametrization singularity described in section $4.4$. Then we arrive the final space $\bar{\check{H}}^{\textrm{s.}}_\gamma:=\check{H}^{\textrm{s.}}_\gamma/\text{Ker}(\Omega_\gamma)$, which is isomorphic by construction to the vertex-simplicity constraint surface in $\mathcal{P}^{\text{S,G}}_\gamma$ as
\begin{equation}
\bar{\check{H}}^{\textrm{s.}}_\gamma/\mathbb{Z}^2\cong \mathcal{P}^{\text{S,G}}_\gamma|_{S_v^{IJKL}=0}.
\end{equation}
after moduling the identifying operation $\mathbb{Z}^2$.
}

\end{itemize}

Based upon this reduction procedure, we claim that the kinematic physical degrees of freedom of the theory on a given graph $\gamma$ are captured by the collection of solutions of verticex-simplicity constraints in the phase space $\bar{\check{H}}_\gamma/\mathbb{Z}^2$.

Let us supply the above reduction procedure in the discrete LQG phase space with a classical picture under proper continuous limits for the Regge sector discussed in Section 3.2, with the interpretation of the variables $(V_e,\tilde{V}_e,N_e,\xi_e,\xi^\mu_e)$ as the extrinsic and intrinsic geometrical data. Specifically, recall that the $N_e$ has the meaning of the area of the (D-1)-face dual to $e$, and $\xi^o_e$ represents the norm of the extrinsic curvature 1-form integrated along $e$ are clear. A concrete translation between $(g_e, X_e)$ and the hypersuface ADM data $(\pi^{aIJ}(x), K_{aIJ}(x))$ can be thus  established using the straight forward conditions $g_e\simeq \mathbb{I}+A_e$ and $X_e\simeq \pi_e$, with which we have
\begin{equation}\label{split1}
\frac{1}{2\beta}\textrm{Tr}(X_edg_eg_e^{-1})\simeq\frac{1}{\beta}\pi^e_{IJ}dA_e^{IJ}.
\end{equation}
In the limit of infinitely short edges one may then read off the sympletic form $\Omega=-d\Theta=\frac{1}{\beta}dA\wedge d\pi$, and the familiar brackets of loop quantum gravity follow,
\begin{equation}
  \{A_{e}^{IJ}(x),\pi^{e'}_{KL}(y)\}=2\beta\delta^{[I}_{K}\delta^{J]}_L\delta_e^{e'}\delta^{(D)}(x-y).
\end{equation}
Also, recalling the splitting
\begin{equation}\label{split2}
A_{a}^{IJ}=\Gamma_a^{IJ}(\pi)+\beta K_a^{IJ}
\end{equation}
with $\Gamma_a^{IJ}(\pi)$ being a function of $\pi^{bKL}$ satisfying $\Gamma_a^{IJ}(\pi)=\Gamma_a^{IJ}(e)$ on simplicity constraint surface, one recovers the Poisson bracket $\{K_{e}^{IJ}(x),\pi^{e'}_{KL}(y)\}=2\delta^{[I}_{K}\delta^{J]}_L\delta_e^{e'}\delta^{(D)}(x-y)$.
The same continuous limit also reveals the classical counterpart to our simplicity constraint reduction in the discrete phase space. Through the correspondence  $g_e=u_ee^{\xi^e\tau_o}e^{\bar{\xi}_e^\mu\bar{\tau}_\mu}\tilde{u}_e^{-1}$ and $u_ee^{\bar{\eta}_e^\mu\bar{\tau}_\mu}e^{\eta^e\tau_o}\tilde{u}_e^{-1}\simeq \mathbb{I}+\Gamma_e$ in the continuous limits being taken, we have
\begin{equation}
K_e  \simeq \frac{1}{\beta}u_e (\xi^o_e\tau_o+\check{\xi}_e^\mu\bar{\tau}_\mu)u_e^{-1},
\end{equation}
with the notation  $e^{\check{\xi}_e^\mu\bar{\tau}_\mu}\equiv e^{-\bar{\eta}_e^\mu\bar{\tau}_\mu}e^{\bar{\xi}_e^\mu\bar{\tau}_\mu}$. Recalling our correspondence $K^{\perp}_e:=\frac{1}{\beta}u_e (\xi^o_e\tau_o)u_e^{-1}$, $K^{/\!/}_e:=\frac{1}{\beta}u_e (\check{\xi}_e^\mu\bar{\tau}_\mu)u_e^{-1}$, we can clearly see that despite of the anomaly in the vertex-simplicity constraints, our reduction procedure correctly remove the component $K^{/\!/}_e$, while preserving the component $K^{\perp}_e$ that contributes to the extrinsic curvature as expressed in the same form as in the classical Ashtekar formulation:
\begin{equation}\label{Kpi}
\text{tr}(K_e\pi^{e'})=\frac{1}{\beta} \text{tr}( u_e (\xi_e^o\tau_o+\check{\xi}_e^\mu\bar{\tau}_\mu)u_e^{-1}\pi^{e'}) =\frac{1}{\beta}\text{tr}( u_e (\xi_e^o\tau_o)u_e^{-1}\pi^{e'})=\text{tr}(K^{\perp}_e\pi^{e'}), \  \  b(e)=b(e').
\end{equation}
Indeed, as a generator of the group preserving $\mathcal{N}_{b(e)}=\mathcal{N}_{b(e')}$, the component $K^{/\!/}_e$ has no projection on the bivector $\pi^{e'}\simeq X^{e'}=N_{e'}V_{e'}=N_{e'}\mathcal{N}^{[I}_{b(e')}V^{J]}_{e'}$ and thus provides no contribution to the extrinsic curvature as it showed in above Eq.\eqref{Kpi}.

This procedure is thus consistent to the symplectic reduction with respect to simplicity constraint in connection phase space in which it act as a well first class constraint, where $K^{/\!/}_e$ play the same role as the component $\bar{K}_{aIJ}$ in connection phase space, and $\text{tr}(K_e\pi^{e'})$ is proportion to the densitized extrinsic curvature $\tilde{K}_a^{\ b}$ along the graph in continuum limit. This result means that we can choose the $SO(D-1)$ fibers as the ``gauge orbit'' of simplicity constraint in the discrete phase space, because the redundant degrees of freedom $K^{/\!/}_e$ are precisely those transformed along the $SO(D-1)$ fibers, which is same as how $\bar{K}_{aIJ}$ acts along the true gauge orbits of simplicity constraint in continuum connection phase space.

Now based on the above discussions, we have demonstrated that
 \begin{itemize}
   \item On both edge and vertex-simplicity constraints surface, the degrees of freedoms in $\bar{\xi}^\mu_e$ (or equivalent, the $SO(D-1)$ fiber) of the discrete LQG phase space represent ``gauge'' degrees of freedom playing the same role of the components $K^{/\!/}$  eliminated in the symplectic reduction with respect to the simplicity constraints in the original Ashtekar formulation of all dimensional LQG in the continuous phase space.
 \end{itemize}
Remakably in this sense, under the correspondence between the generalized twisted geometry variables and smeared Ashtekar variables in the Regge sector, the continuous limit of our reduction procedure indeed recovers the symplectic reduction in the Ashtekar formulation with respect to the original Gauss and simplicity constraints.

\section{Conclusion and outlook}

To better explore the spacetime geometry information encoded in the higher-dimensional spin-network states, we proposed a new kinematic gauge-reduction procedure for the $SO(D+1)$ LQG at the classical and discrete level. The reduction takes place with respect to the anomalous kinematic constraint system consisting of the discrete simplicity constraints $ \{ S_e^{IJKL}\approx0,\,S_v^{IJKL}\approx0 \}$ and the discrete $SO(D+1)$ Gauss constraints $G_v^{IJ}\approx0$, defined in the $SO(D+1)$ LQG phase space associated with a given graph for the spin network states.

Motivated by our previous work on the weak solutions of the quantum vertex-simplicity constraints given by the coherent intertwiners, we generalized the twisted-geometry parametrization of the $SU(2)$ LQG phase space, into the angle-bivector parametrization of the constraint surface of edge-simplicity constraints $S_e^{IJKL}\approx0$ in the $SO(D+1)$ LQG phase space. Further, when restricted to the common constraint surface of the full kinematic constraints $ \{ G_v^{IJ}\approx0,\,S_e^{IJKL}\approx0,\,S_v^{IJKL}\approx0 \}$, the new parametrization endows the angle-bivector variables with the meaning of the constrained smeared formulations of Ashtekar variables. In particular, the $SO(D-1)$ angle variables are identified with the smeared Ashtekar connection components that are pure-gauge corresponding to simplicity constraint in the original classical and continuous theory.

Through studying the properties of the Hopf sections in $SO(D+1)$ Hopf fibre bundle, we obtained the Poisson algebra among the angle-bivector variables, and subsequently the action of the constraint system on the  twisted-geometry variables. Then, the full symplectic reduction with respect to the first-class sub-system of the discrete constraints $\{ G_v^{IJ}\approx0,\, S_e^{IJKL}\approx0\}$ is performed and results to the gauge-invariant reduced phase space $\mathcal{P}^{\text{S,G}}_\gamma$. Crucially as we discovered, when again restricted to the common constraint surface of the full system of the discrete constraints, the first-class subsystem generates the orbits that recover the gauge orbits generated by the original continuous simplicity and Gauss constraints in the continuous limits. In particular, the edge-simplicity constraints generate precisely the transformations in the $SO(D-1)$ angle variables. Finally, we demonstrated that when the remaining anomalous vertex-simplicity constraints are imposed as additional constraints upon the gauge-invariant reduced phase space, the selected state space $ \bar{\check{H}}^{\textrm{s.}}_\gamma\subset \mathcal{P}^{\text{S,G}}_\gamma $ truly describes the discrete ADM data in the form of Regge hypersurface geometry, up to the shape matching condition.

We are thus led to the new point of view, in which the (quantum) discrete Gauss and edge-simplicity constraints are the true generators of the kinematic gauge symmetry for all-dimensional LQG , while the anomalous vertex-simplicity constraints only act as addition state-selection laws.
From a different view, this new gauge reduction has supplemented the missing pieces in our previous prescription of the weak solutions to the quantum vertex-simplicity constraint, where the weak solutions are interpreted as semi-classical polytopes used to assemble the spatial geometry. Indeed, the absent description of the extrinsic curvature components in the Ashtekar connection are captured by the angle variables parametrizing the holonomies, left out from the previous prescriptions based on only the bivectors labelling coherent intertwiners. Remarkably, our gauge orbit reductions leave the only angle variable $ \xi^o_e $ giving precisely the (smeared) Regge extrinsic curvature, for the states on the vertex-simplicity constraint surface in the reduced phase space.

Our results point to two interesting future research directions based upon the generalized twisted-geometry parametrization of the $SO(D+1)$ LQG phase space. Firstly, we have assumed the existence of the coherent spin-network states sharply peaked in the angle-bivector variables--- this was based on the known explicit construction of the $SU(2)$ coherent spin-network states sharply peaked in the twisted-geometry variables \cite{Bianchi:2009ky}. As indicated in this earlier study, these states are highly valuable for not only do they recover the boundary semi-classical states for the spinfoam models (a covariant version of LQG), but also serve as a special type of the Thiemann-Hall's complexifier coherent states in canonical LQG \cite{Thiemann_2001}\cite{Thiemann_20012}\cite{Thiemann_20013}. Through these connections, the clear intrinsic and extrinsic geometry interpretations via the twisted-geometry variables have illuminated many important perspectives of the classical limits for both canonical and covariant LQG in $(1+3)$ dimension. In the same manner, we expect our angle-bivector parametrization to offer valuable insights to the covariant and canonical LQG in higher dimensions, for which the geometric meaning of the coherent states has been even more elusive. In fact, it is known that the currently prevailing Thiemann-Hall's $SO(D+1)$ coherent states are too complicated for explicit computations. It is our hope that the angle-bivector coherent states, which could be constructed and studied based on the recent works \cite{long2020perelomov}, may serve as the alternative coherent states with the much simpler Gaussian distribution formulation and clear geometric meanings, for clarifying the semi-classical behavior of the $SO(D+1)$ LQG. The second direction is toward the physical evolutions in canonical $SO(D+1)$ LQG. This may be pursuit either in the context of the Dirac theory with the Hamiltonian operators (arbitrary combinations of the quantum scalar and vector constraints) to be solved as additional quantum constraints and the local observables to be constructed, or in the context of a classically deparametrized theory with one physical Hamiltonian operator giving the evolution in a specified notion of time. In both cases, the new crucial challenge here is to deal with the algebra involving the quantum Hamiltonian operators. As mentioned, although the full system of Gauss, simplicity and Hamiltonian constraints are of first class in the continuous classical theory, it inevitably becomes anomalous under the loop quantization \cite{bodendorfer2013newiii} \cite{Gaoping2019geometricoperators}--- especially with the typical closed loop holonomy representation for the curvature factors in the Hamiltonian operators. Our new insights in the quantum orbits for the simplicity constraints may provide an approach to the problem that is closely guided by the physical and geometric picture. For instance in the context of the deparametrized theory, our results suggest quantizing the discretized physical Hamiltonian associated to each graph that is gauge invariant with respect to just the quantum Gaussian constraints and edge-simplicity constraints. This would lead to the dynamics preserving the gauge symmetry, with which one could then study the weak stability of the vertex-simplicity constraints under the dynamics. Indeed, the program in this manner would be guided by the ultimate goal for the quantum evolutions of the Regge ADM data.

\section*{Acknowledgments}
This work is supported by the National Natural Science Foundation of China (NSFC) with Grants No. 11775082, No. 11875006 and No. 11961131013.

\bibliographystyle{unsrt}

\bibliography{ref}

\end{document}